\documentclass[preprint,pre,aps,showkeys,nofootinbib,amsmath,amssymb,floatfix]{revtex4-2}
\usepackage{amsmath}
\usepackage{amssymb}
\usepackage{graphicx}
\usepackage{dcolumn}
\usepackage{bm}
\usepackage{float}
\usepackage{placeins}
\usepackage{subcaption}

\begin{document}

\preprint{APS/123-QED}

\title{Coupling-Induced Synchronized Motion and Stochastic Resonance in Overdamped Dimers}

\author{Dhruv Agrawal}
 \altaffiliation[Also at ]{Department of Physics, National Institute of Technology Meghalaya}
\author{W. L. Reenbohn }%
 \email{wlreenbohn@nitm.ac.in}
\affiliation{%
 Department of Physics, National Institute of Technology Meghalaya, Shillong, Meghalaya 793003, India
}%



\date{\today}

\begin{abstract}
In this study, we explore an overdamped system of a dimer in a bistable potential immersed in a heat bath. The monomers interact via the combination of the Lennard-Jones potential and the harmonic potential. We have introduced a short-range interaction in our model making it more physical. Such a classical system can be used as a model for stochastic resonance (SR) based energy harvesters where the interplay between the noise, coupling and a periodic perturbation leads to a rich class of dynamical behaviours. A key distinction between observing SR in single and coupled particle studies is that a transition between the two wells is only considered successful if both the particles cross a certain threshold position. Although we observe qualitatively a similar peaking behaviour in different quantifiers of SR (like input energy ($W_p$) and hysteresis loop area (HLA)), the effects of the above-mentioned condition on the dynamics of the system remain unaddressed to the best of our knowledge. We study SR using different measures like the input energy per period of the external forcing, the hysteresis loop area as well as quantities like phase lag between the response and the external forcing and the maximum average amplitude of the response. Additionally, we have defined a new quantity called the successful transition ratio. This ratio helps us understand the effects of the dimer's coupling on the number of successful transitions out of the total attempted transitions. The successful transition ratio is almost unity for strongly coupled dimer suggesting most of the transition attempts end up successfully however few they are in numbers. On the other hand, the ratio shows a peaking behaviour with respect to noise for weak and intermediate couplings. We show that only for the weakly coupled dimer, the ratio is maximum around the temperature where SR takes place.
\end{abstract}

\keywords{Stochastic resonance, bistable system, coupled nonlinear system, input energy, hysteresis loop area.}

\maketitle

\section{\label{sec1}Introduction \protect}

\begin{figure}[H]    
    \includegraphics[width=1\textwidth]{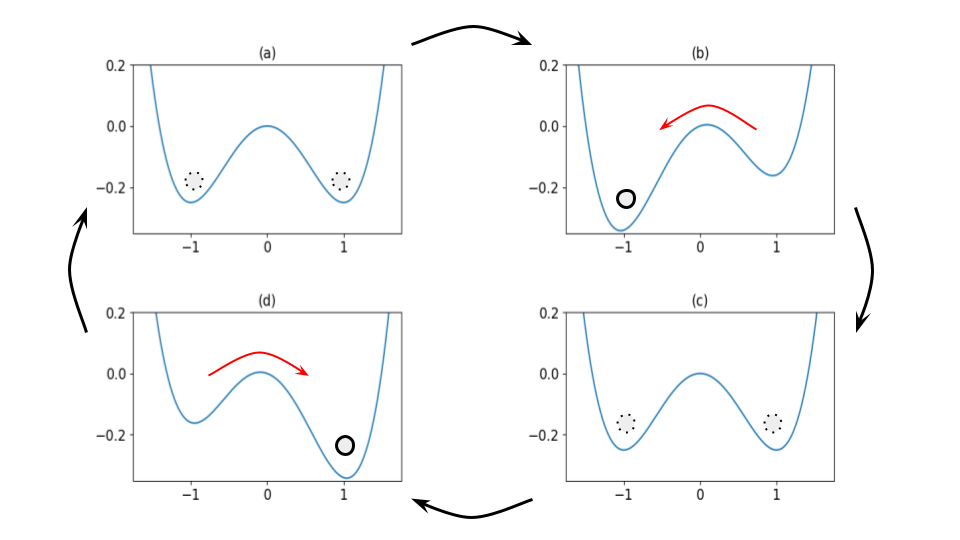}
    \caption{The phenomena of SR}
    \label{fig:suitcnd}
\end{figure}

Stochastic resonance (SR) refers to the phenomena wherein the response of a nonlinear system to a weak periodic force is optimized due to the presence of noise in the system \cite{wiesenfeld1995stochastic,gammaitoni1998stochastic}. This results in an additional periodicity in the output which is due to the complex interplay between the noise, the external periodic force and the non-linearity of the system. Introduced by Benzi et al. \cite{benzi1981mechanism,benzi1982stochastic,benzi1983theory} to explain the 100 thousand year cycle of glacial and inter-glacial periods in the Earth's climate, SR has since then been subsequently investigated in various fields of physics, neurology, and engineering. Some of the notable studies include the study of SR in a Schmitt trigger \cite{fauve1983stochastic}, a $\mu
m$ sized Brownian particle \cite{simon1992escape}, the mechanoreceptor hair cells of a crayfish \cite{douglass1993noise}, different kinds of neuronal models \cite{plesser1997stochastic, lu2008dynamics, xu2020dynamics}, medical image-processing \cite{peng2007stochastic}, mechanical fault detection \cite{qiao2019applications} and sustainable energy harvesting \cite{zheng2014application}. In all these areas, the noise plays a constructive role by detecting and amplifying weak signals or by aiding the transitions between the states in the system. 
\\SR has been studied in systems exhibiting multiple stable states \cite{reenbohn2012periodically,xu2018stochastic,xu2019stochastic, mei2021characterizing, xu2023study} out of which the bistable systems are the most commonly studied systems. When a particle is kept in one of the stable states of such a bistable potential, it makes random transitions between the stable states due to the noise present in the system. On introducing an external periodic force and tuning the system parameters carefully, the periodicity of transitions between the states can be synchronised with the external periodic force. SR occurs when the timescale at which the particle makes a transition between the wells (that depends upon the noise strength subjected to the system) matches with half the period of the external periodic force \cite{benzi1983theory,gammaitoni1989stochastic,gammaitoni1998stochastic}. At this moment, the external drive forces one of the stable states of the bistable potential to become relatively less stable than the other. In other words, the transition from the less stable state to the more stable state becomes more probable than the reverse transition. Such a condition is favourable for the particle to make a transition from the less stable to the more stable state. After half a period of the external drive, the stability of the states reverses. Fig. \ref{fig:suitcnd} shows this change in the stability of the states of the potential over one period of the external drive. If the noise enables the particle to make exactly one transition across the barrier during this favourable condition, the particle is able to show the same periodicity as the external drive and the response of the system gets enhanced. This is the phenomena of SR explained in the most general system of a single particle kept in a bistable potential. However, the systems occurring in nature are generally multi-particle systems having different forms of coupling between them. The rapid developments in science and technology, especially in the field of energy harvesting has led to an increase in theoretical and experimental studies on complex coupled systems. The interaction between the coupled particles introduces new dimensions to the resonance phenomenon. The system 

\noindent of two coupled particles i.e. a dimer, is the simplest example of a coupled multi-particle system, providing us a rich platform to investigate how coupling influences the noise assisted phenomena. Several interesting aspects are observed when the coupling between the particles is taken into account. For example, a much more efficient collective SR effect is obtained when a periodic force is added to a system of globally coupled particles under the influence of noise \cite{jung1992collective}, the enhancement of information flow is obtained in a system of N globally coupled nonlinear oscillators under the influence of noise and a periodic force, where one of the oscillators relaxes at a much smaller time scale as compared to the others \cite{bulsara1993stochastic}, the existence of an optimum value of coupling responsible for an enhanced collective response of a two coupled bistable system \cite{neiman1995stochastic}, vibrational and stochastic resonance was studied in two coupled overdamped anharmonic oscillators in \cite{gandhimathi2006vibrational,gandhimathi2007vibrational}, an experimental study focusing on frequency entrainment and synchronization in coupled nanomechanical oscillators \cite{shim2007synchronized}, two driven underdamped bistable systems under the influence of independent noises were studied and SR was controlled by changing the relative phase between the two driving forces involved in \cite{kenfack2010stochastic}. More recent works include the application of SR techniques in coupled systems to detect faults in rolling bearings \cite{li2019novel} and in the domain of energy harvesting \cite{aravindan2022array}. The latter shows that for a fixed coupling strength and noise level, the harvested power saturates for a specific system size. Such works motivate this study to investigate SR in complex coupled systems. To the best of our knowledge, there has not been much work studying SR for a dimer having the Lennard-Jones (LJ) interaction potential as a part of its coupling potential. We believe that including the LJ potential makes the model more robust by taking into account the fact that any two coupled sub-structures (molecules or oscillators) should not collide with each other. They should not even come close to each other beyond a certain critical distance. There needs to be a short range repulsive force that would prevent these sub-structures from closing in and collapsing upon themselves. The LJ potential is commonly used in molecular physics and chemistry to model interaction between a pair of neutral atoms or molecules. It provides an effective model to replicate the interaction between particles. The potential includes a repulsive term which corresponds to the Pauli's exclusion principle leading to a short range repulsion between two or more particles, and an attractive term modelling weak attractive forces at longer distances. The LJ component and the harmonic interaction together can be considered as the coupling for the dimer. This coupling introduces new resonant conditions which are now sensitive to more than just the noise strength and the external periodic force. In the following sections, we aim to examine how the coupling parameters influence the synchronization in the transitions made by the dimer, when the system is subjected to an external periodic force. We also explore the rich dynamics of a dimer and the mechanisms by which SR manifests itself in the system.

The paper is structured in the following manner: Section \ref{sec2} provides the details about our model, section \ref{sec3} discusses the numerical results obtained based on the simulations performed to explore the escape dynamics of the dimer, the phenomena of SR quantified using HLA and input energy and section \ref{sec4} includes the discussion and conclusions.

\section{\label{sec2}The model of an overdamped dimer kept in a 1-dimensional bistable potential \protect}
We explore an overdamped dimer kept in a 1-dimensional bistable potential $V(x)$ given by, 
\begin{equation}\label{BP}
    V(x) = -\frac{\omega_B^2}{2}x^2 + \frac{\omega_B^2}{4x_m^2}x^4
\end{equation}
where, $\pm x_m$ denotes the position of the minima of the potential, the barrier height $V_B=x_m^2\omega_B^2/4$ and $\omega_B^2=\omega_0^2/2$. $\omega_B^2$ and $\omega_0^2$ are the curvature of the potential at the top of the barrier and the bottom of the potential well respectively.

The dimer has 2 unit mass monomers coupled together by a spring of equilibrium length $l_0$, having a spring constant $k$. Hooke's force $F_H$ is responsible for the restoring force between the monomers. The Lennard-Jones (LJ) force $F_{LJ}$ acting between them provides a repulsive force that prevents them from collapsing upon each other. Eqns. (\ref{hp}-\ref{lp}) give the Hooke's and the LJ potential respectively.

\begin{equation}\label{hp}
    V_H(r) = \frac{1}{2}k(r - l_0)^2
\end{equation}

\begin{equation}\label{lp}
    V_{LJ}(r) = 4\epsilon\left(\left( \frac{\sigma}{r}\right)^{12}-\left(\frac{\sigma}{r}\right)^{6} \right)
\end{equation}
where $r = \lvert x_1-x_2 \rvert$, gives the length of the dimer, $\epsilon$ represents the depth of the potential and $\sigma$ is the length of the dimer corresponding to which the LJ potential vanishes. It is related to the equilibrium length of the dimer as,
\begin{equation}
    \sigma = \frac{l_0}{2^{\frac{1}{6}}}
\end{equation}
The corresponding forces $F_H$ and $F_{LJ}$ are,
\begin{equation}
    F_H = -k(r-l_0)
\end{equation}
\begin{equation}
    F_{LJ} = 4\epsilon\left(12\left( \frac{\sigma^{12}}{r^{13}}\right)-6\left(\frac{\sigma^6}{r^7}\right) \right)
\end{equation} 
The system is subjected to a heat bath at some constant temperature T. This effect is modelled as the Gaussian white noise $\eta(t)$ having the following properties,

\begin{equation}
    \langle \eta_i(t)\rangle = 0
\end{equation}

\begin{equation}
    \langle\eta_i(t)\eta_j(t')\rangle = 2D\gamma\delta_{ij}\delta(t-t')
\end{equation}
If the coefficient of friction is $\gamma$ and the noise strength is $D$, the dynamics of the dimer can be represented by the following coupled Langevin equations,

\begin{align}
    \gamma\frac{\mathrm{d}x_1}{\mathrm{d}t} &= -V'(x_1) + F_H(r)+ F_{LJ}(r) + \eta_1(t) \nonumber \\
    & = x_1 - x_1^3 + F_H(r)+ F_{LJ}(r) + \eta_1(t)
\end{align}

\begin{align}
    \gamma\frac{\mathrm{d}x_2}{\mathrm{d}t} &= -V'(x_2) - F_H(r)- F_{LJ}(r) + \eta_2(t) \nonumber \\
    & = x_2 - x_2^3 - F_H(r)- F_{LJ}(r) + \eta_2(t)
\end{align}
An external weak sinusoidal force of magnitude $A_0$ and angular frequency $\Omega$ is added to the system. The Langevin equations now become,
\begin{equation}\label{eq:fn1}
     \gamma\frac{\mathrm{d}x_1}{\mathrm{d}t} = x_1 - x_1^3 + F_H(r)+ F_{LJ}(r) +A_0\sin{\Omega t}+ \eta_1(t)
\end{equation}

\begin{equation}\label{eq:fn2}
    \gamma\frac{\mathrm{d}x_2}{\mathrm{d}t} = x_2 - x_2^3 - F_H(r)- F_{LJ}(r) +A_0\sin{\Omega t}+ \eta_2(t)
\end{equation}

For numerical simplicity we have non-dimensionalized eqns. (\ref{eq:fn1}-\ref{eq:fn2}) by introducing dimensionless time $\bar{t}=t/\tau$, length $\bar{x}_i=x_i/x_m$, coupling strength $\bar{k}=k/\omega_B^2$, depth of the potential $\bar{\epsilon}=\epsilon/V_B$, equilibrium length $\bar{l}_0=l_0/x_m$, noise strength $\bar{D}=D/V_B$, Gaussian white noise $\bar{\eta_i}=\eta_i(t)x_m/V_B$, amplitude of the external drive $\bar{A}_0=A_0x_m/V_B$ and its frequency $\bar{\Omega}=\Omega\tau$. The constant $\tau$ is calculated to be $\gamma x_m^2/V_B$. The dimensionless coupled Langevin equations are,\footnote{The bar over each dimensionless quantity is ignored for simplicity.}

\begin{equation}\label{dlfn1}
     \frac{\mathrm{d}x_1}{\mathrm{d}t} = 4(x_1 - x_1^3) -4k(r-l_0) + 4\epsilon\left(12\left( \frac{\sigma^{12}}{r^{13}}\right)-6\left(\frac{\sigma^6}{r^7}\right) \right) +A_0\sin{\Omega t}+ \eta_1(t)
\end{equation}

\begin{equation}\label{eq:dlfn2}
    \frac{\mathrm{d}x_2}{\mathrm{d}t} = 4(x_2 - x_2^3) + 4k(r-l_0) - 4\epsilon\left(12\left( \frac{\sigma^{12}}{r^{13}}\right)-6\left(\frac{\sigma^6}{r^7}\right) \right) +A_0\sin{\Omega t}+ \eta_2(t)
\end{equation}
The properties of Gaussian white noise in dimensionless units are,
\begin{equation}
    \langle \eta_i(t)\rangle = 0
\end{equation}

\begin{equation}
    \langle\eta_i(t)\eta_j(t')\rangle = 2D\delta_{ij}\delta(t-t')
\end{equation}

\section{\label{sec3}Numerical results \protect}
Three regimes of coupling between the individual units of the dimer are considered i.e. a soft dimer $(k=0.05)$, a hard dimer $(k=1.0)$ and a dimer having an intermediate value of coupling strength $(k=0.2)$. SR is investigated by studying the variation of several quantifiers with the noise strength and the coupling strength. The data has been recorded for the positions of the monomers as well as their center of mass and averaged over a total of 100 ensembles. The value of parameters used in the numerical simulation are:
\begin{enumerate}
    \item $l_0 = 0.4$
    \item $\epsilon = 0.1$
    \item $A_0 = 0.1$
    \item $t = 10^6$
    \item $dt = 10^{-3}$
    \item $\Omega = 0.018$
\end{enumerate}

The data of different trajectories is plotted with the corresponding values of the external drive to obtain the hysteresis loops. Hysteresis loops are trajectories in the average position-force $(\langle{x}\rangle-F)$ space. The area enclosed within these loops (HLA) corresponds to the energy lost by the system to the environment per period of the external drive. According to \cite{evstigneev2005quantifying}, HLA characterizes the work done by the system over each cycle of the external driving force. Apart from providing knowledge about the energy exchanged, the hysteresis loops also provide information about the average maximum amplitude of the output signal and the phase relationship between the output signal and the external drive.

For the calculation of input energy and HLA per period, we have divided the data obtained from the trajectories in segments corresponding to every 1 period of the external drive. The input energy per period is defined as the energy pumped into the system of the dimer per period by the external drive. In terms of work done, the input energy is equivalent to average work done per period $(W_p)$ on the system by the external drive. From \cite{sekimoto_stochastic_2010}, we get the expression for the input energy as,
\begin{equation}
    W_p = \int_{t_0}^{t_0+t_\Omega}\frac{\partial U(x_1,x_2,t)}{\partial t}\, dt
\end{equation}
The term $U(x_1,x_2,t)$\footnote{From eqns. (\ref{dlfn1}-\ref{eq:dlfn2}) it is evident that $U(x_1,x_2,t)$ will have a slightly different form for $x_1$ and $x_2$. Since the part of the potential which is independent of time gets eventually eliminated due to differentiation, we have not written the expressions corresponding to $x_1$ and $x_2$ separately.} includes the contribution due to the bistable, LJ, Hooke's as well as external periodic force. We can simplify it by expressing $U(x_1,x_2,t)$ as,
\begin{equation}
    U(x_1,x_2,t)=U(x_1,x_2)-xA_0\sin \Omega t
\end{equation}
Therefore, the expression for input energy becomes,
\begin{equation}
    W_p = -A\Omega\int_{t_0}^{t_0+t_\Omega}x\cos{\Omega t}\, dt
\end{equation}

With this insight, we discuss the different quantifiers analysed in our study in the upcoming sections. We start by analyzing the hysteresis loops corresponding to the different coupling strengths of the dimer. The hysteresis loops provide crucial information for other physical quantities like average maximum amplitude ($\langle x_A \rangle$) and phase lag ($\phi$) that supplement our study. We explore how these quantities change over time while the noise strength, the coupling and the frequency of the external drive are adjusted in a controlled manner.

\subsection{\label{sec:hla}Hysteresis loops \protect}
\begin{figure}[H]
    \centering
    \begin{subfigure}[b]{0.42\textwidth}
        \centering
        \includegraphics[width=\textwidth]{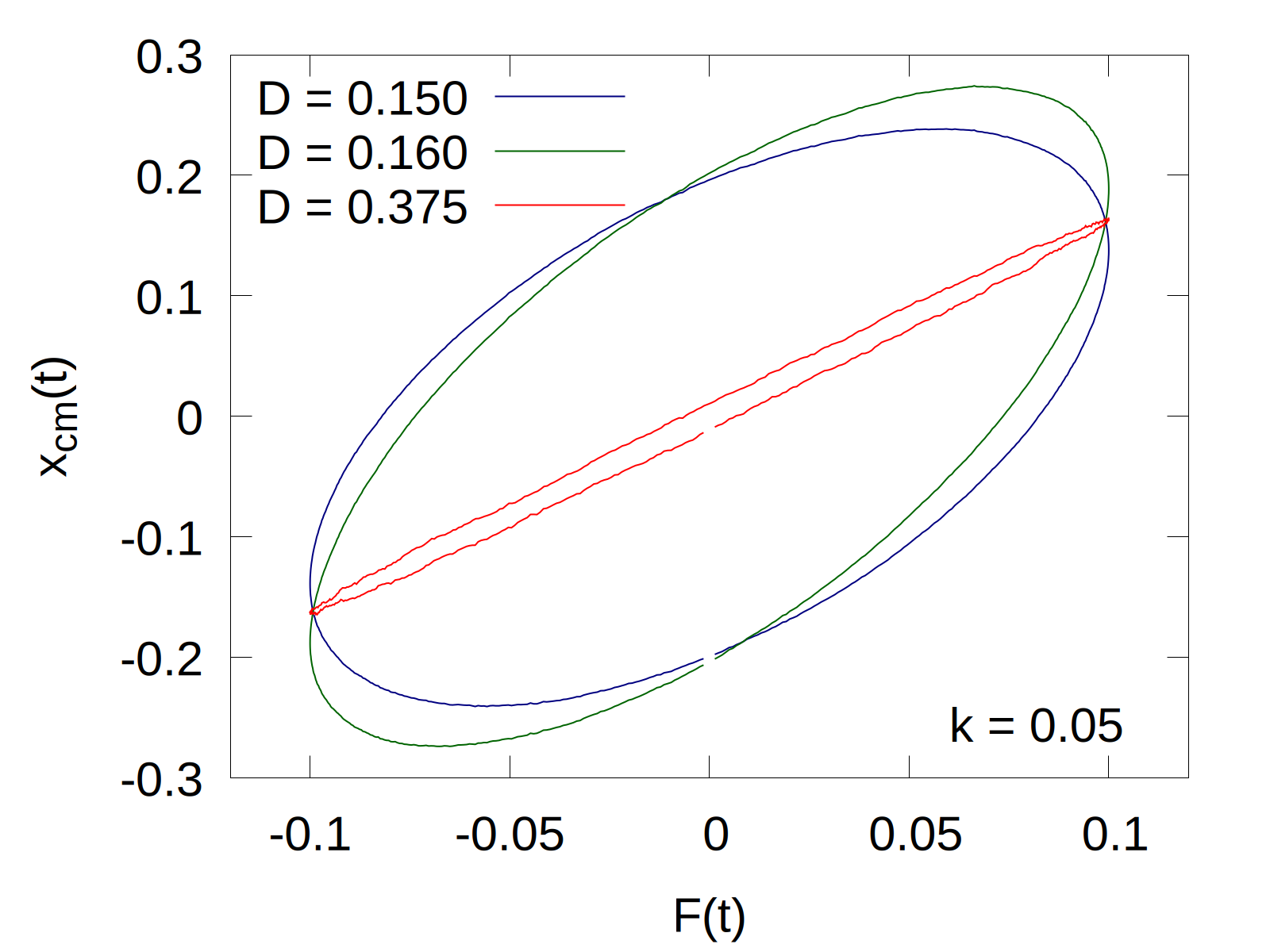}
        \caption{$k=0.05$}
        \label{fig:hlimg1}
    \end{subfigure}
    \hfill
    \begin{subfigure}[b]{0.42\textwidth}
        \centering
        \includegraphics[width=\textwidth]{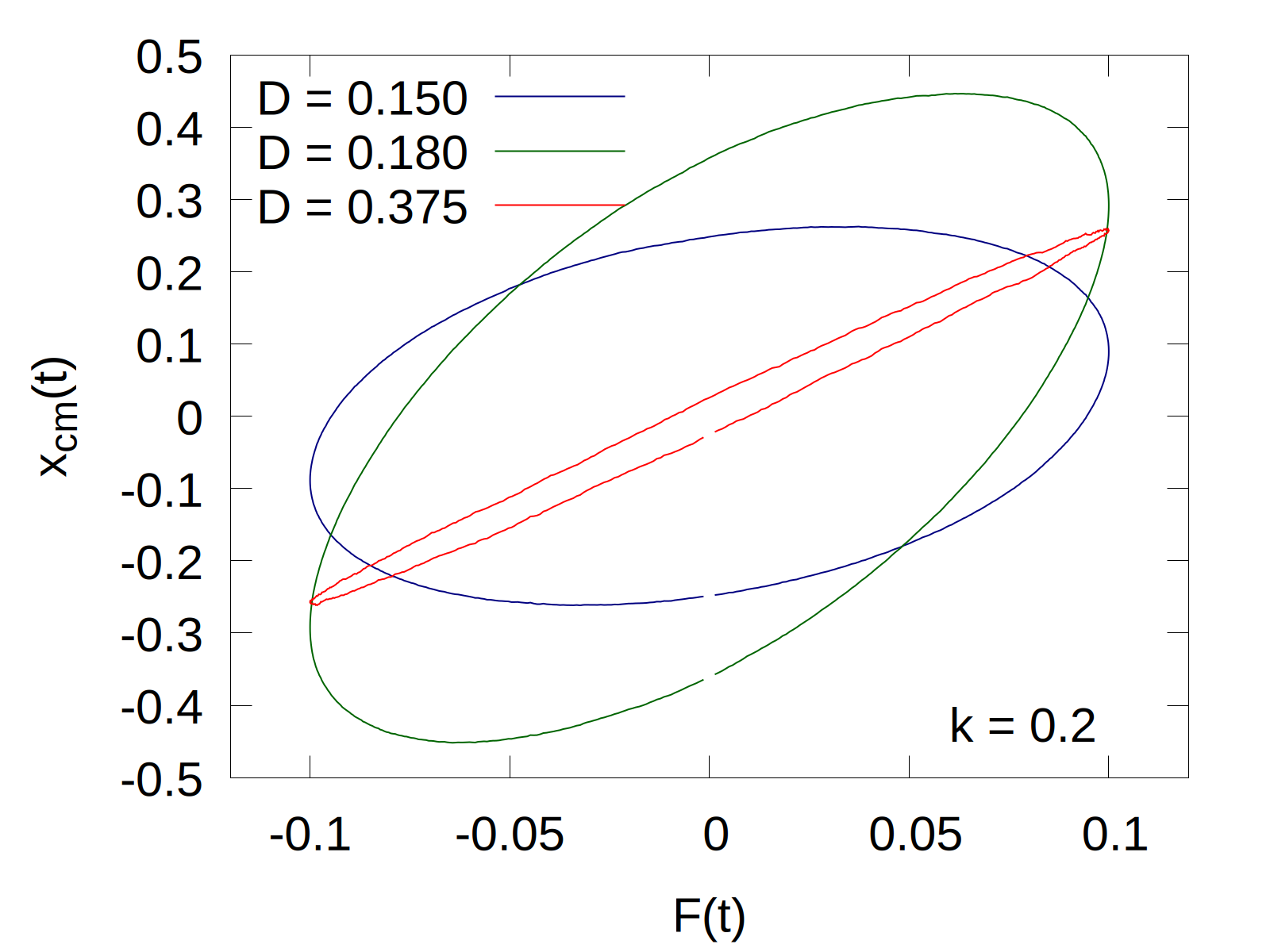}
        \caption{$k=0.2$}
        \label{fig:hlimg2}
    \end{subfigure}
    \hfill
    \begin{subfigure}[b]{0.42\textwidth}
        \centering
        \includegraphics[width=\textwidth]{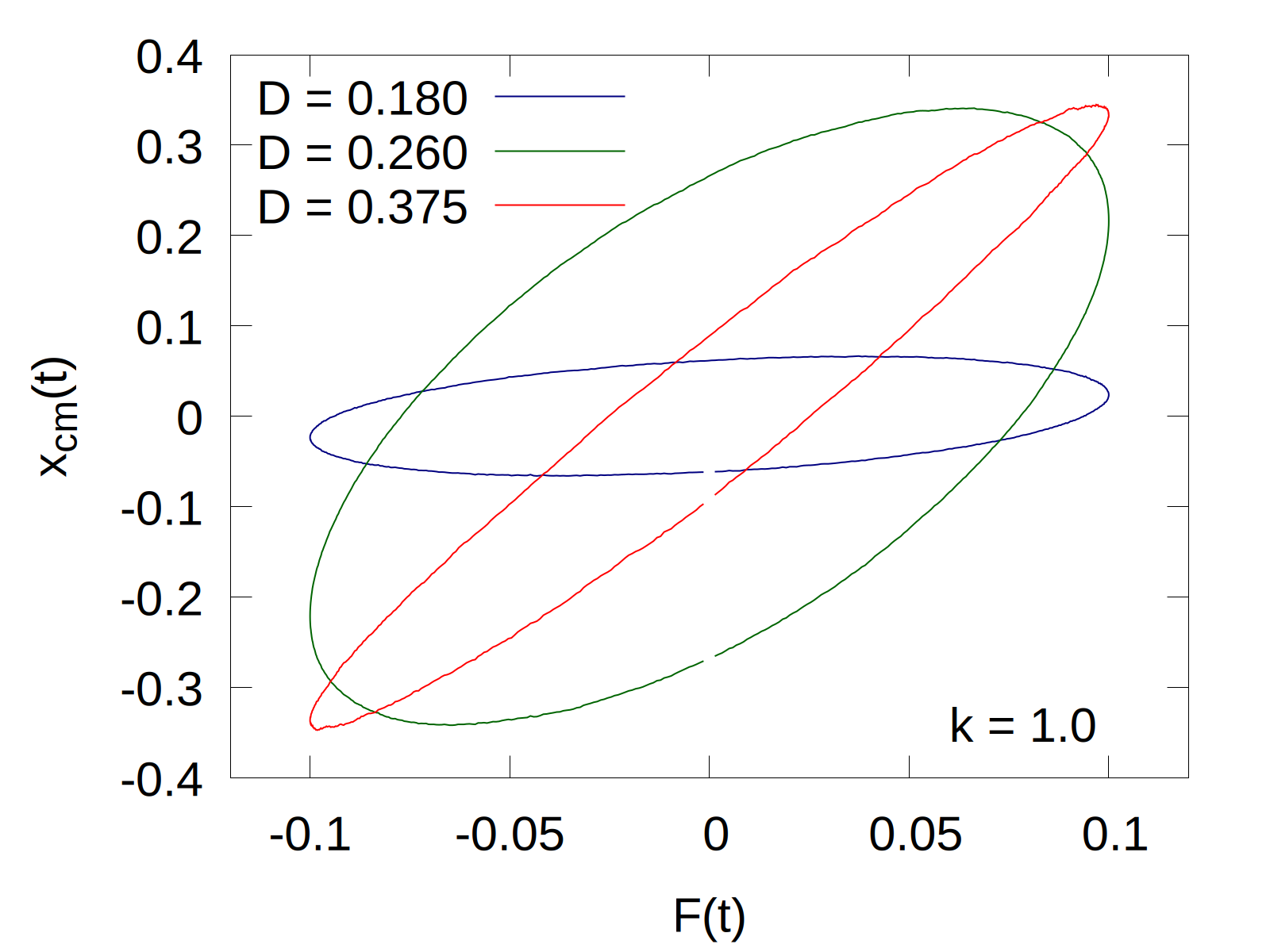}
        \caption{$k=1.0$}
        \label{fig:hlimg3}
    \end{subfigure}
    \caption{Hysteresis loops for $x_{cm}$ of the 3 regimes of the dimer i.e. (a) $k=0.05$, (b) $k=0.2$ and (c) $k=1.0$. The green loop in each plot corresponds to the value of D for which SR is observed.}
    \label{fig:hlthree_images}
\end{figure}

\begin{figure}[H]
    \centering
    \begin{subfigure}[b]{0.44\textwidth}
        \centering
        \includegraphics[width=\textwidth]{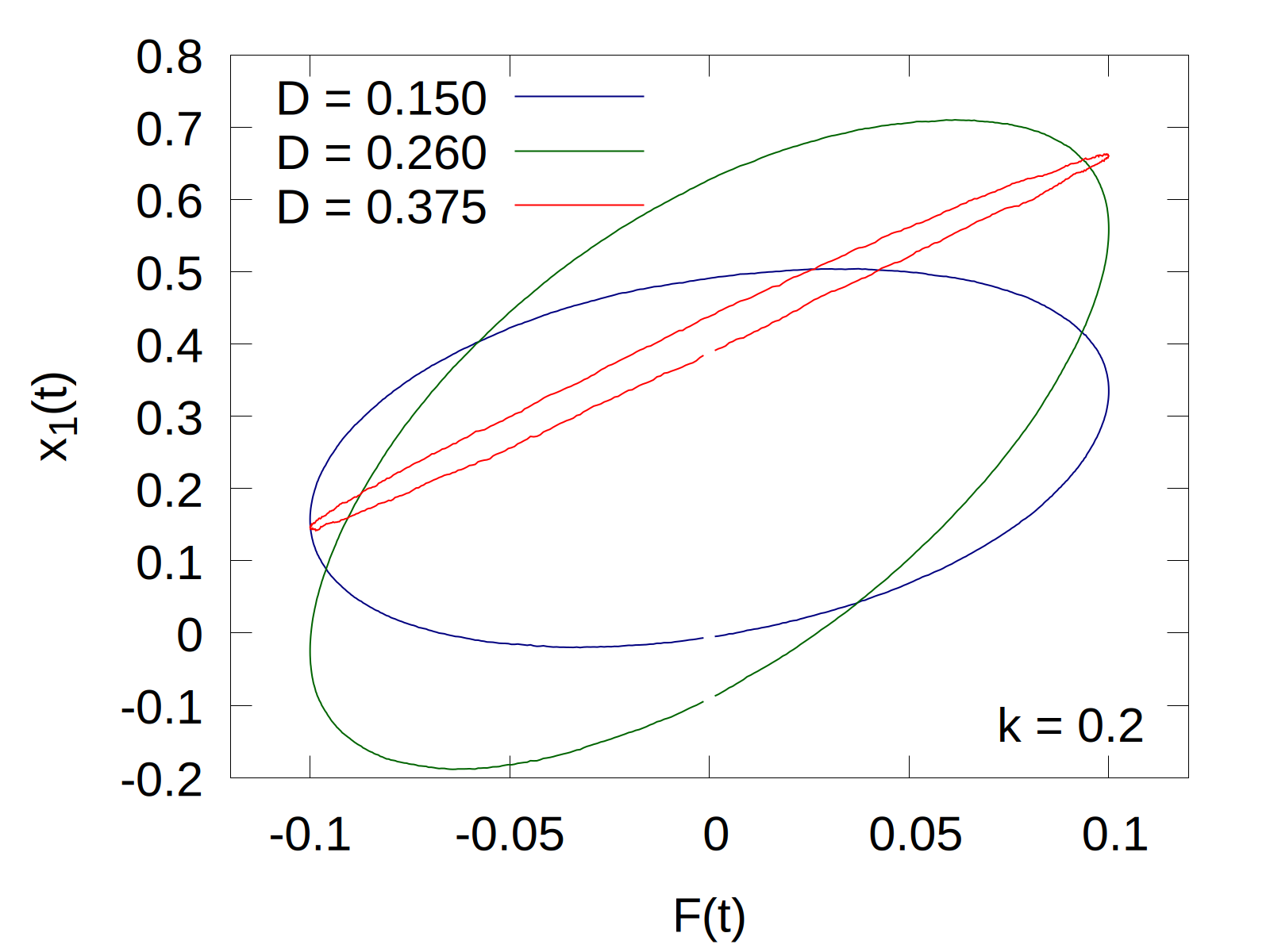}
        \caption{$x_1$}
        \label{fig:hl1img1}
    \end{subfigure}
    \hfill
    \begin{subfigure}[b]{0.44\textwidth}
        \centering
        \includegraphics[width=\textwidth]{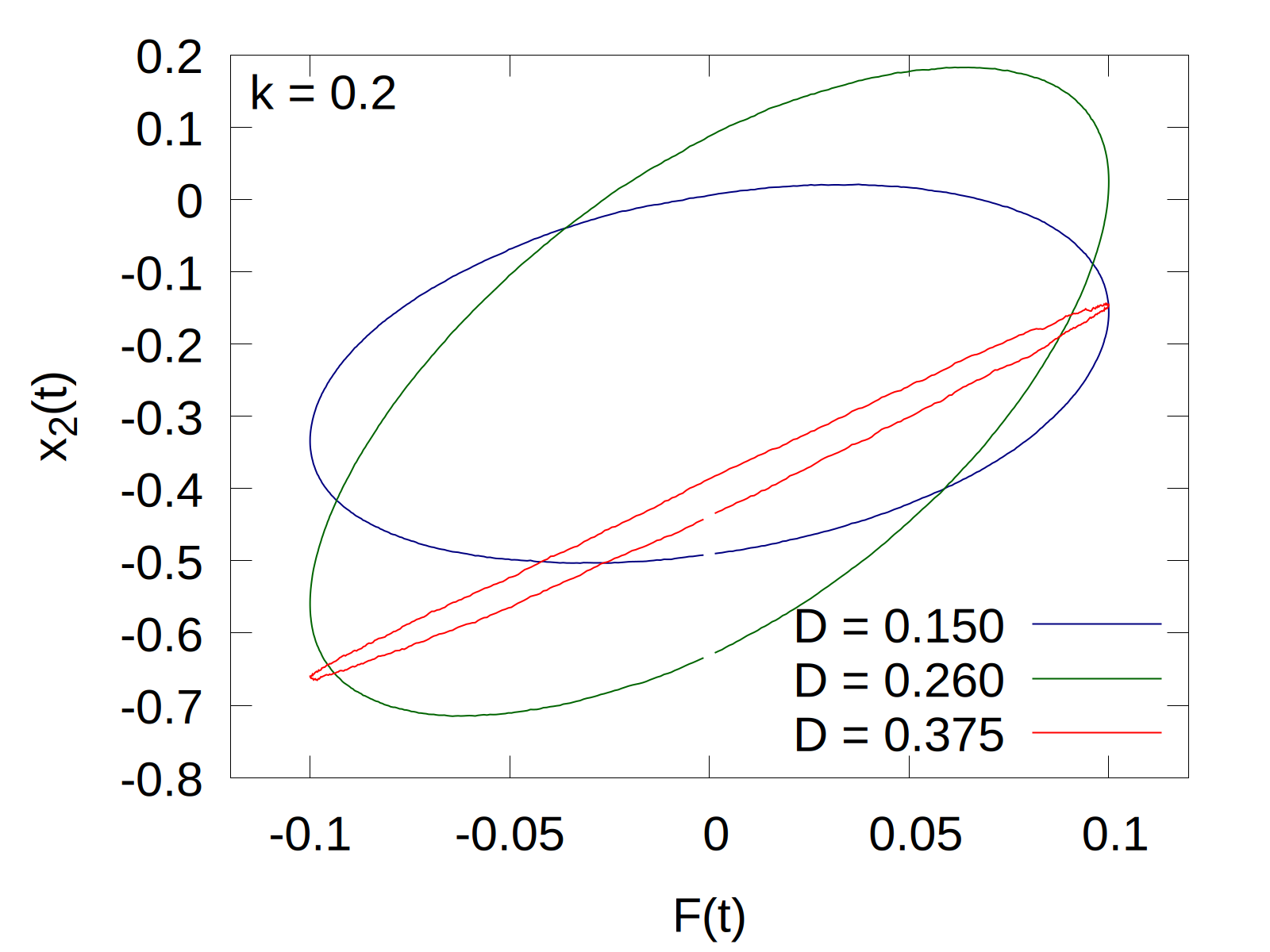}
        \caption{$x_2$}
        \label{fig:hl2img2}
    \end{subfigure}
    
    \caption{Comparison of Hysteresis loops for $x_{1}$ and $x_{2}$ of the dimer having $k=0.2$. The green loop in each plot corresponds to the value of D for which SR is observed.}
    \label{fig:hl122three_images}
\end{figure}

The hysteresis loops are curves in the $(\langle{x}\rangle-F)$ space that are used to gain an insight on how the average position of the center of mass of the dimer ($\langle{x}_{cm}\rangle$) responds to the applied weak periodic perturbation $F(t)$. The loop characterises the lag between the external force and the response of the system due to the complex interplay between the bistable potential, thermal fluctuations and the periodic force.
The energy pumped into the system per period by the external drive (input energy) or the average work done by the external drive on the system is dissipated by the system due to damping over one cycle. The energy dissipated per cycle is represented by the area of the hysteresis loop. The plots in Fig. \ref{fig:hlthree_images} show the hysteresis loops for the center of mass of the dimer, for the three coupling regimes at different noise intensities. 
The loop with maximum area corresponds to the condition of SR because the dimer dissipates maximum energy during SR. For noise levels ($D$), below the one corresponding to SR ($D_{SR}$) i.e. ($D<D_{SR}$), the dimer fails to efficiently absorb energy from the external drive. This leads to smaller loop areas. On increasing the noise strength, the dimer starts absorbing and hence dissipating more and more energy leading to an increase in the HLA. At an optimum value of noise i.e. $D_{SR}$, the transitions of the dimer become coherent with the external drive and hence, the energy input is maximised. This also maximises the energy dissipated. On further increasing the noise strength, the fluctuations start dominating the dynamics of the dimer and the motion of its center of mass becomes less coherent with the external drive. The most noticeable difference for the three noise levels can be seen for a hard dimer in Fig. \ref{fig:hlimg3}.
 \noindent The hysteresis loops for the position coordinate of the monomers have also been plotted in Fig. \ref{fig:hl122three_images}. The plots show the change in the hysteresis loop area for a dimer having an intermediate coupling strength for three different noise strengths. The position of the hysteresis loops differ from those recorded in Fig. \ref{fig:hlthree_images} as the position of the center of mass of the dimer is the average of the position of the monomers. A strong repulsive LJ force keeps the monomers apart. The average position of the monomer to the left is skewed towards smaller values as compared to the position of the monomer on the right. As the response of the system is different at different noise strengths, the hysteresis loops for the monomers get shifted and rotated accordingly. This will also be reflected in the phase lag analysis of the dimer. The single particle studies in a bistable potential show a similar rotation of the ellipse with the changing noise strength but, we do not observe any such shifting of the loops. This shift can be attributed to the presence of an LJ interaction between the monomers. The ordinate of the center of the ellipse corresponding to the average position of the monomers takes different values due to the finite equilibrium length of the dimer and the fluctuations about it.

Next, we analyse the trajectories of the center of mass coordinate of the dimer that are averaged over 100 ensembles for different coupling coupling strengths and temperature. 
\subsection{Ensemble averaged trajectories and average maximum amplitude}
\begin{figure}[H]
    \centering
    \includegraphics[width=0.75\textwidth]{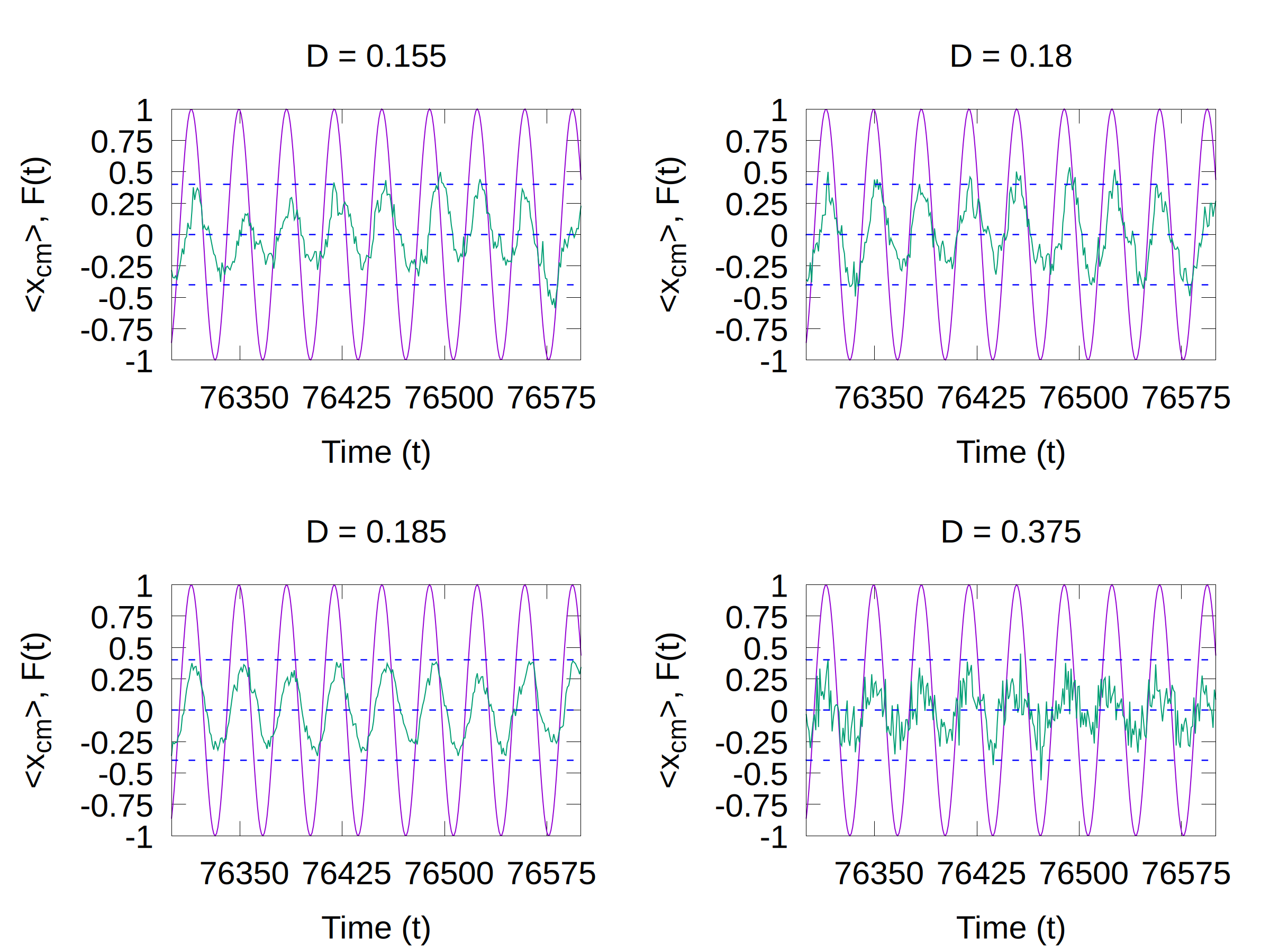}
    \caption{Ensemble averaged trajectories of center of mass of the dimer having coupling strength $k=0.05$}
    \label{fig:xcm0.05}
\end{figure}

\begin{figure}[H]
    \centering
    \includegraphics[width=0.75\textwidth]{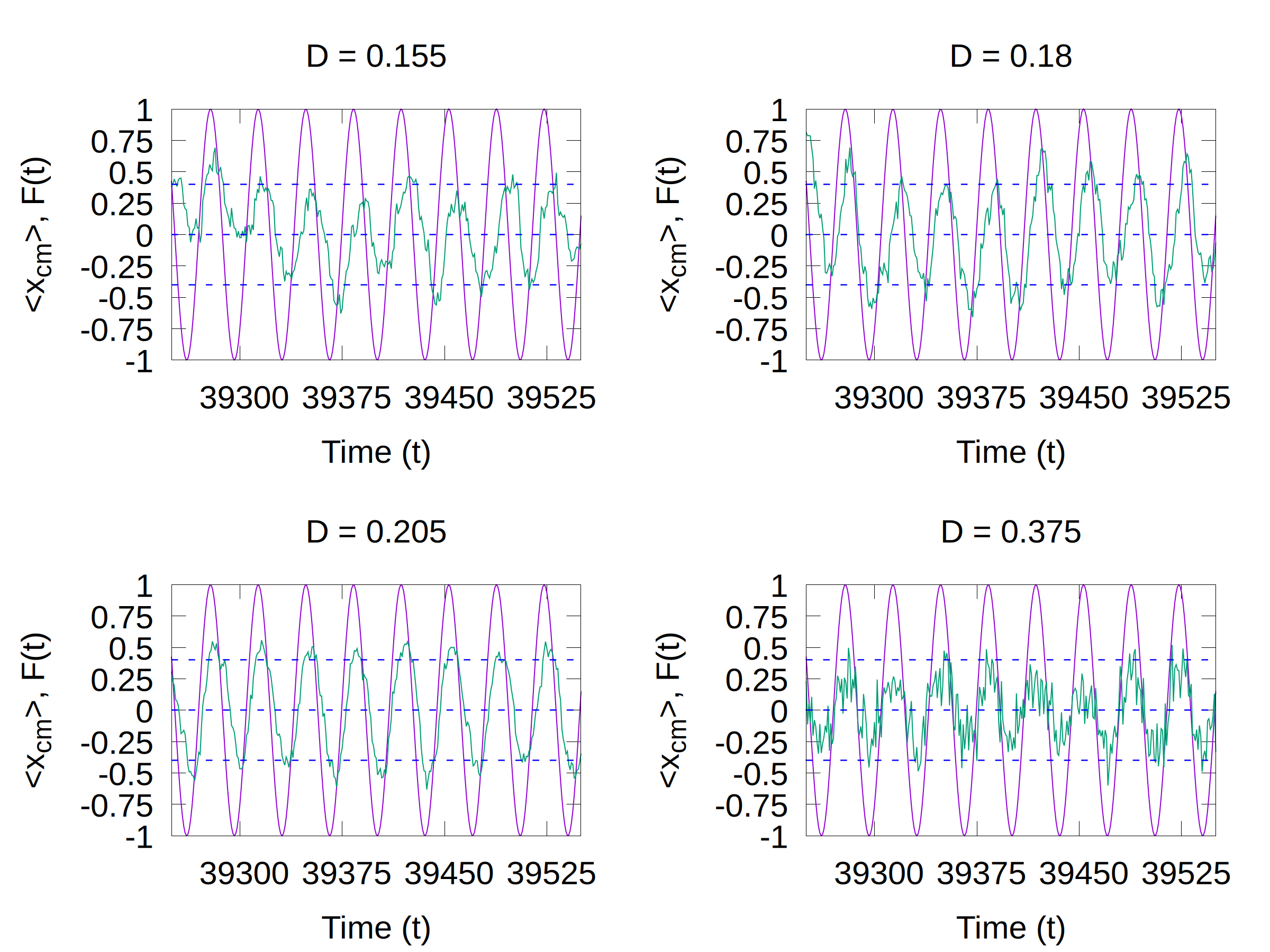}
    \caption{Ensemble averaged trajectories of center of mass of the dimer having coupling strength $k=0.2$}
    \label{fig:xcm0.2}
\end{figure}

\begin{figure}[H]
    \centering
    \includegraphics[width=0.75\textwidth]{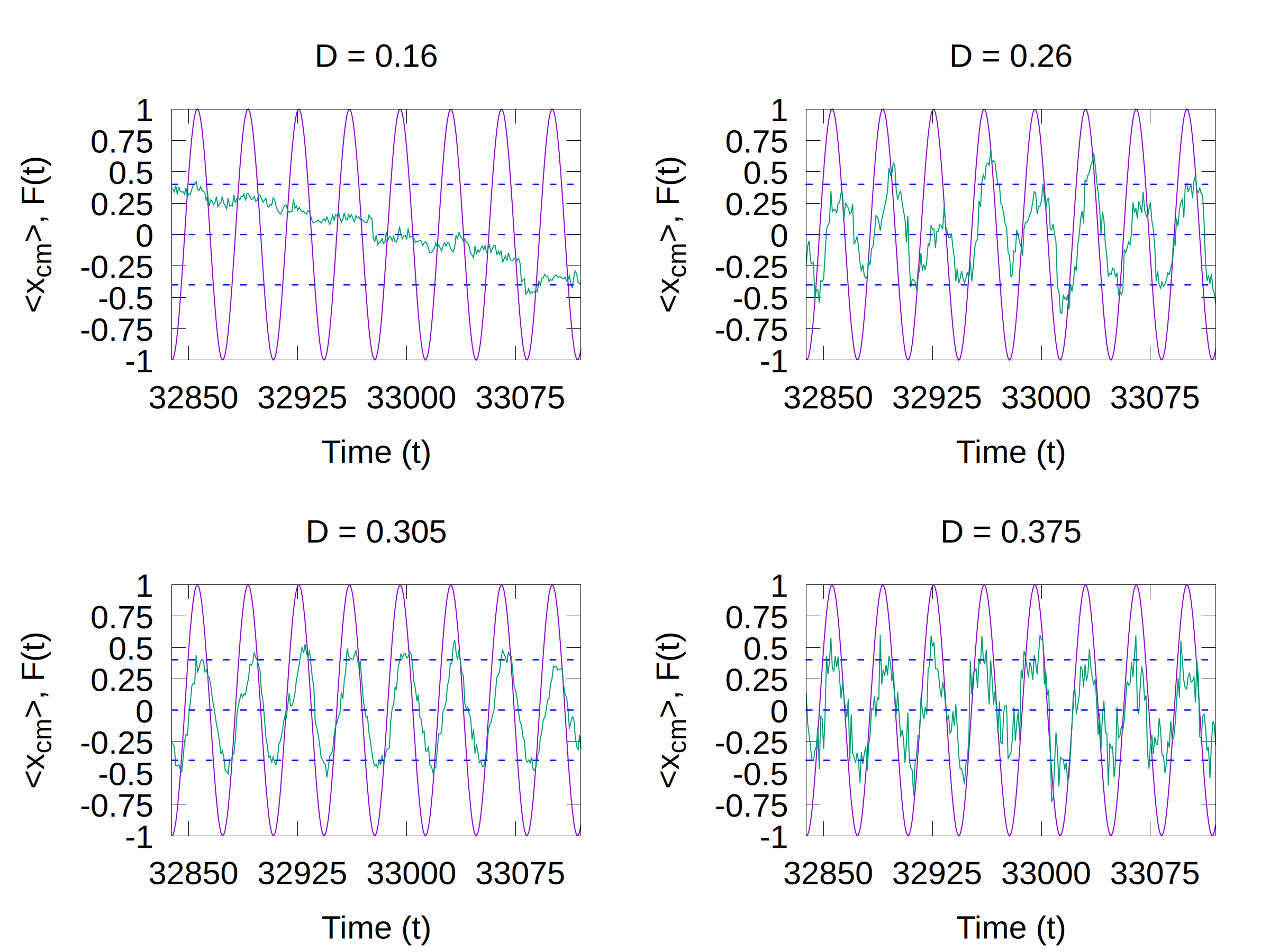}
    \caption{Ensemble averaged trajectories of center of mass of the dimer having coupling strength $k=1.0$}
    \label{fig:xcm1.0}
\end{figure}

The trajectories for the center of mass of the dimer have been analysed with respect to time. First, we have kept the coupling strength of the dimer and the frequency of the external drive fixed. Similar to the case of a single particle in an overdamped bistable system \cite{ray2006stochastic}, we observed a peaking behavior in the average maximum amplitude for $\langle x
_{cm}\rangle$ at some optimum value of noise strength as an indication of SR. Small noise strengths correspond to the case where the thermal energy is not sufficient to make the dimer cross the potential barrier even once per period of the external drive. This results in small average maximum amplitudes obtained in  trajectories of the center of the mass of the dimer. The average maximum amplitudes increase to an approximate value of $0.5$. The dimers having a coupling strength $k=0.05, 0.2$ and $1.0$ show an optimum response at noise strengths $D=0.185, 0.205$ and $0.305$ respectively. Fig. \ref{fig:xcm0.05}-\ref{fig:xcm1.0} highlight an important feature that makes the trajectories of $\langle x_{cm}\rangle$ at their corresponding $D_{SR}$ different from those at other noise strengths. Along with a maximum in average maximum amplitude, the trajectories show the least degree of fluctuations. The increase in fluctuations is a direct result of an increase in the random nature of the dynamics and the number of times the dimer makes a jump across the barrier in a single period. The individual monomers that are weakly coupled can be located in separate wells for a considerably larger part of the trajectory, leading to smaller values of the center of mass. On increasing the temperature, when their escape rate matches with $\Omega$, it will be more likely that the they make a transition simultaneously. Thus increasing

\begin{figure}[H]
    \centering
    \begin{subfigure}[b]{0.43\textwidth}
        \centering
        \includegraphics[width=\textwidth]{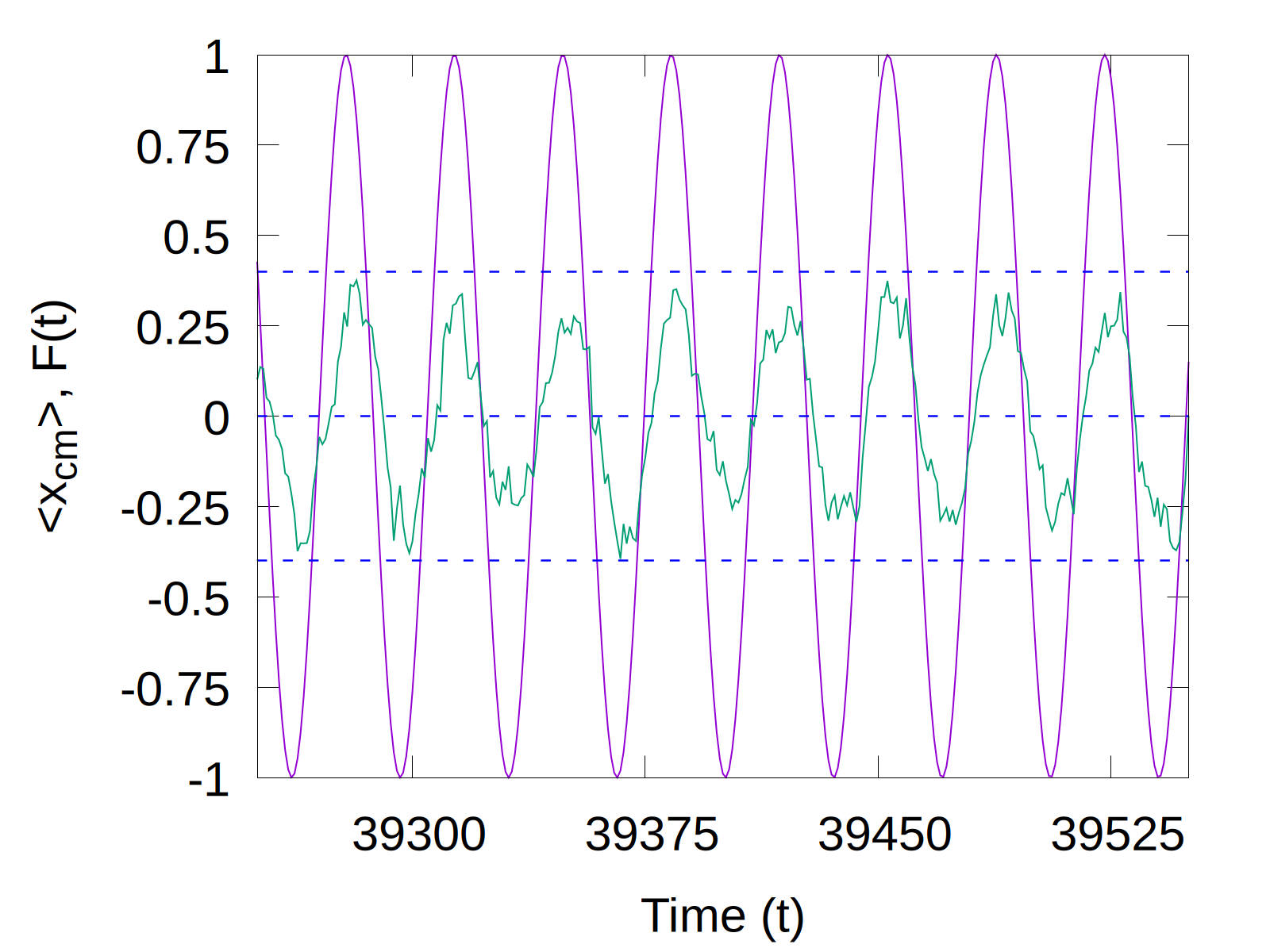}
        \caption{$k=0.05$}
        \label{fig:enavtrk_0.05}
    \end{subfigure}
    \begin{subfigure}[b]{0.43\textwidth}
        \centering
        \includegraphics[width=\textwidth]{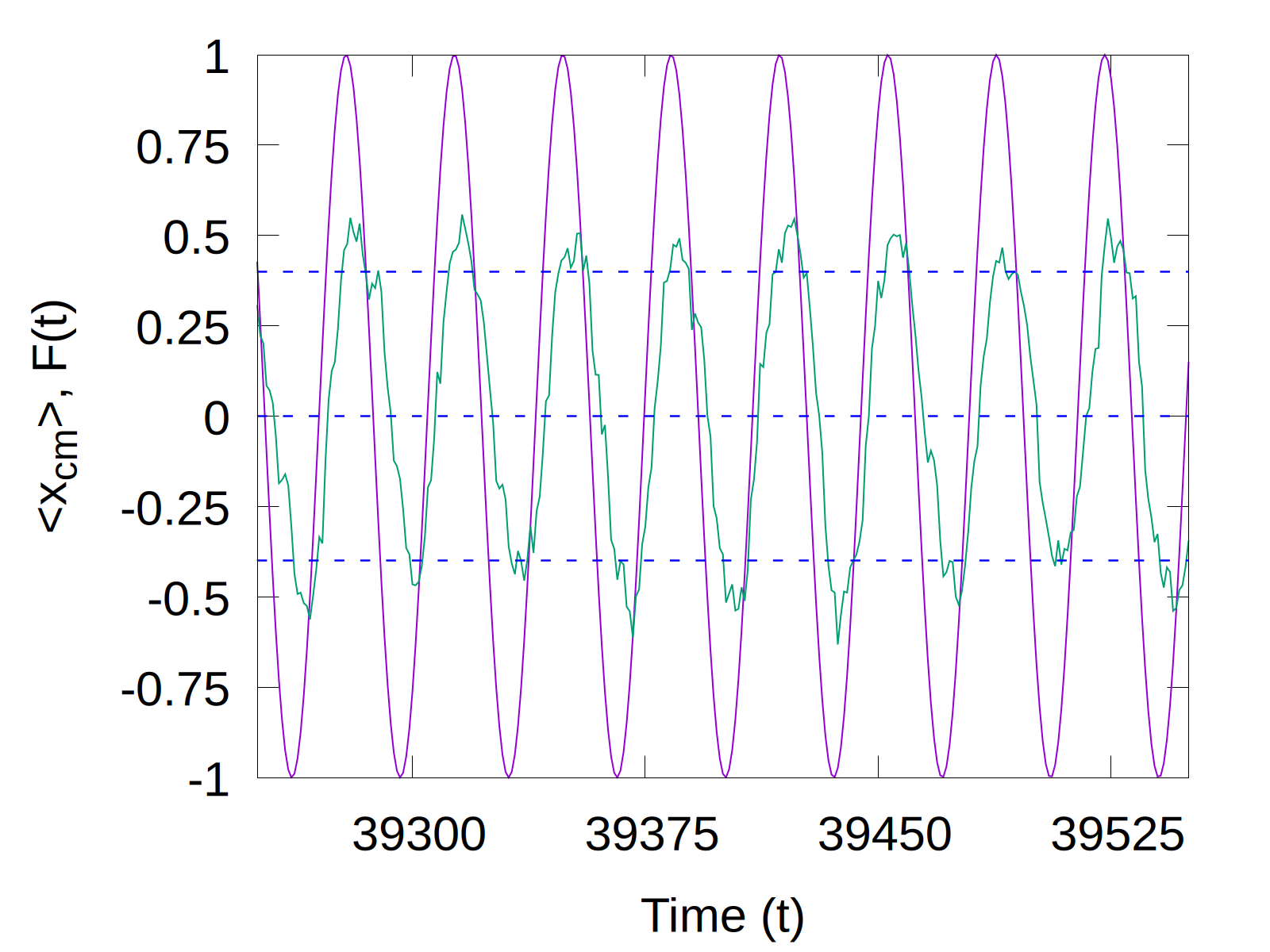}
        \caption{$k=0.2$}
        \label{fig:enavtrk_0.2}
    \end{subfigure}
    \begin{subfigure}[b]{0.43\textwidth}
        \centering
        \includegraphics[width=\textwidth]{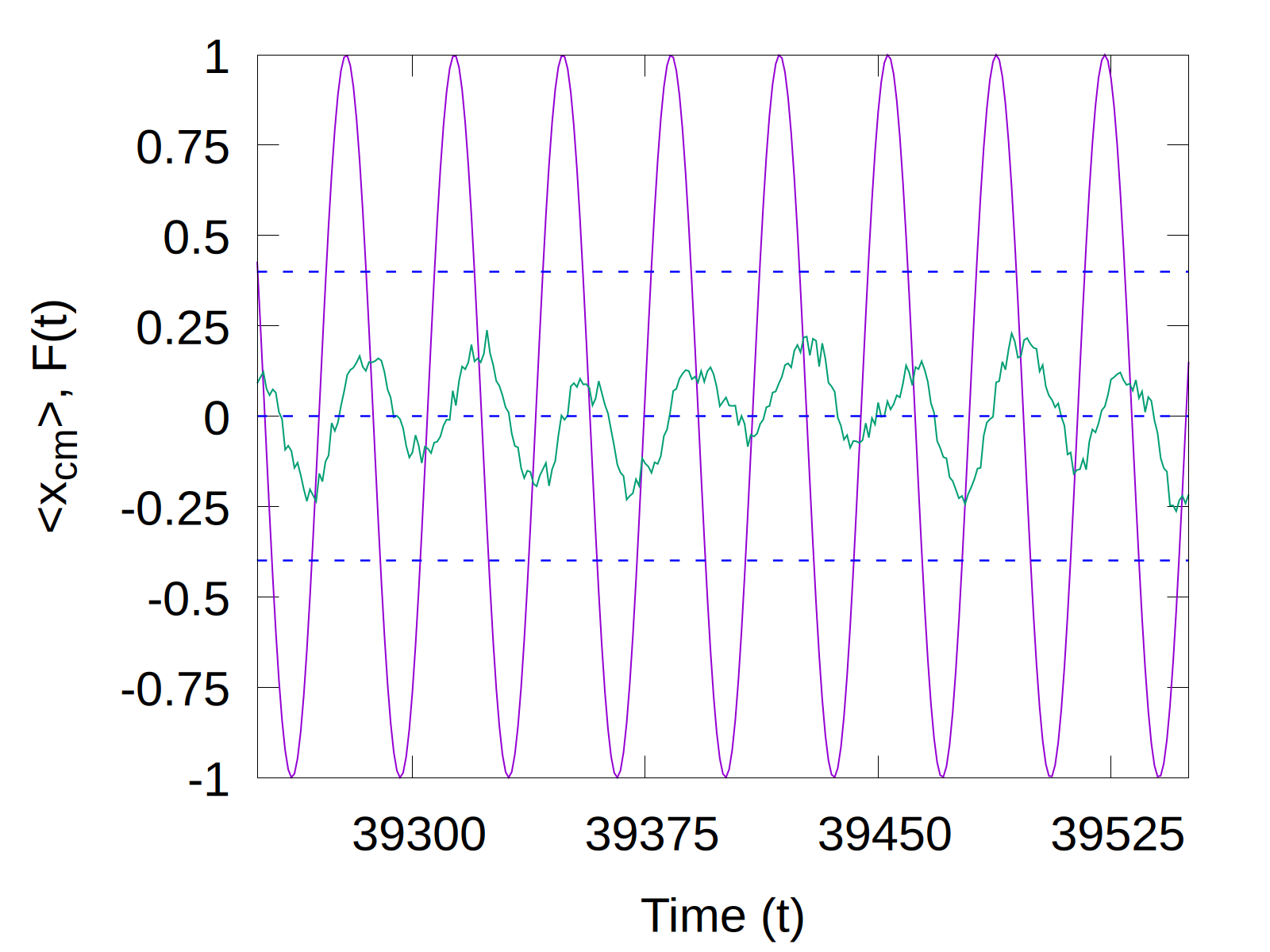}
        \caption{$k=1.0$}
        \label{fig:enavtrk_1.0}
    \end{subfigure}
    \caption{Ensemble averaged trajectories of center of mass of the dimer at $D=0.205$ (a) $k=0.05$, (b) $k=0.2$ and (c) $k=1.0$.}
    \label{fig:xcmd0.18}
\end{figure}

\begin{figure}[H]
    \centering
    \includegraphics[width=0.6\textwidth]{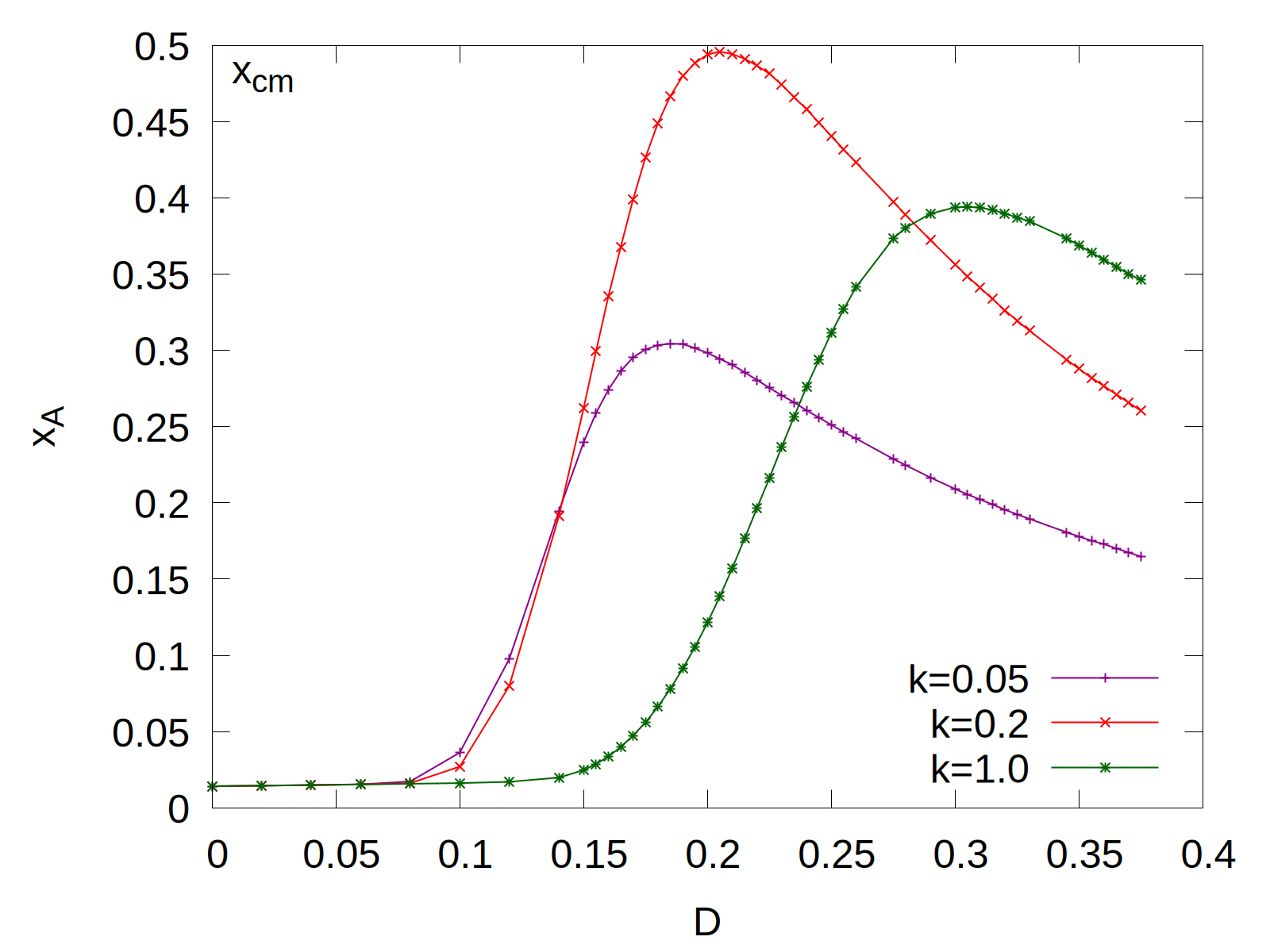}
    \caption{Comparison of average maximum amplitudes of $x_{cm}$ for different coupling strengths of the dimer.}
    \label{fig:mxcmp}
\end{figure}

\noindent their probability of being together in the same well. This leads to an increase in the average amplitudes at an optimum temperature. On further increasing the temperature, again the transitions occur independently for both the monomers along with an increase in their relative distance. This will cause the fluctuations in the average values of the center of mass to increase. For intermediate and strong coupling regimes, the probability that the monomers will follow each other during an attempted transition event increases. This results in higher amplitudes in the average trajectories as compared to the weak coupling case, as observed in Fig. \ref{fig:mxcmp}. At the optimum temperature, this movement becomes more likely to occur and hence we obtain smoother trajectories. 
Fig. \ref{fig:xcmd0.18} shows trajectories of average maximum amplitude of the center of mass of the dimer at a fixed noise strength, i.e. $D=0.205$ for all the three coupling regimes.

\noindent As we increase the coupling strength, there is an optimum value for which the dimer transitions back and forth across the barrier exactly once per period of the external drive and we obtain a maximum average amplitude for the center of mass of the dimer. This is because the monomers essentially don't feel a strong restoring force which would restrict their motion. In order to attain the condition of SR, their must be 1 transition of the dimer across the barrier per half a period of the external drive. In order to achieve that, when the coupling strength is slightly increased in the regime of intermediate coupling, the restoring forces gradually become stronger and are able to restrict the motion of the dimer such that they make at most one transition per half a period. We observe the maximum in the amplitude of the trajectory for intermediate values of $k$. On further increasing the coupling strength, the restoring forces become strong enough to reduce the number of transitions per period across the barrier to less than one. This is indicated by smaller amplitudes in the trajectory shown in Fig. \ref{fig:enavtrk_1.0}.

Fig. \ref{fig:mxcmp} shows a comparison between the average maximum amplitudes ($\langle x_A \rangle$) of the center of mass for the different coupling strengths of the dimer. 
The amplitude tends to show a peaking behaviour even when the coupling strength of the dimer is varied. The temperature at which the amplitude takes a maximum shifts to larger values as the coupling increases. This happens because, with an increase in the coupling strength, the restoring forces start restricting the independent motion of the monomers. Consequently, larger noise strengths are required for the dimer to overcome these restoring effects in order to make a transition across the barrier. In the plot for $x_A$ vs $D$, we observe that the peak is highest for intermediate coupling strength of the dimer. This is due to the synchronization effects and the flexibility of the dimer. For weak coupling, the dynamics of the individual monomers is more independent. This leads to less efficient synchronized movement due to coupling and hence a lower amplitude. When the coupling is strong, the movement of the dimer is constrained. The strong coupling strength restricts the dimer's ability to stretch leading to slightly reduced amplitudes. It is only for the intermediate values of the coupling that the dimer is able to maintain a balance between the synchronized motion due to coupling and its ability to stretch. This increases the overall amplitude at the SR peak for intermediate values of coupling.

Another crucial characteristic of the system, measured from the hysteresis loops is the phase lag $\phi$ between the dimer's motion and the external periodic force. It provides an important insight into the delay in the dimer's response to the periodic force. The following section describes this phase relationship of the dimer in detail.
\subsection{Phase lag\protect}\label{Section:pl}
\begin{figure}[H]
    \centering
    \begin{subfigure}[b]{0.4\textwidth}
        \centering
        \includegraphics[width=\textwidth]{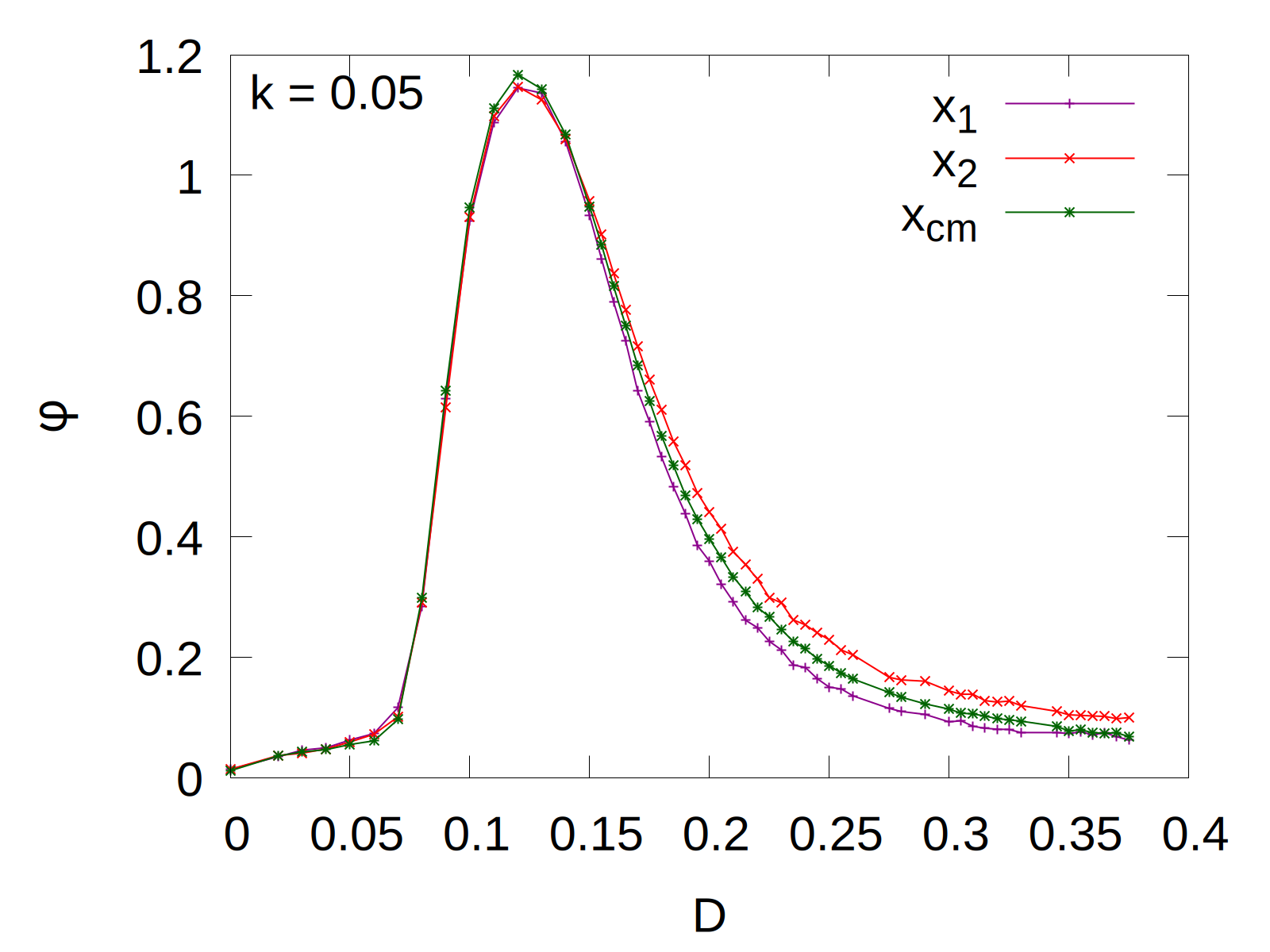}
        \caption{$k=0.05$}
        \label{fig:plimg1}
    \end{subfigure}
    \hfill
    \begin{subfigure}[b]{0.4\textwidth}
        \centering
        \includegraphics[width=\textwidth]{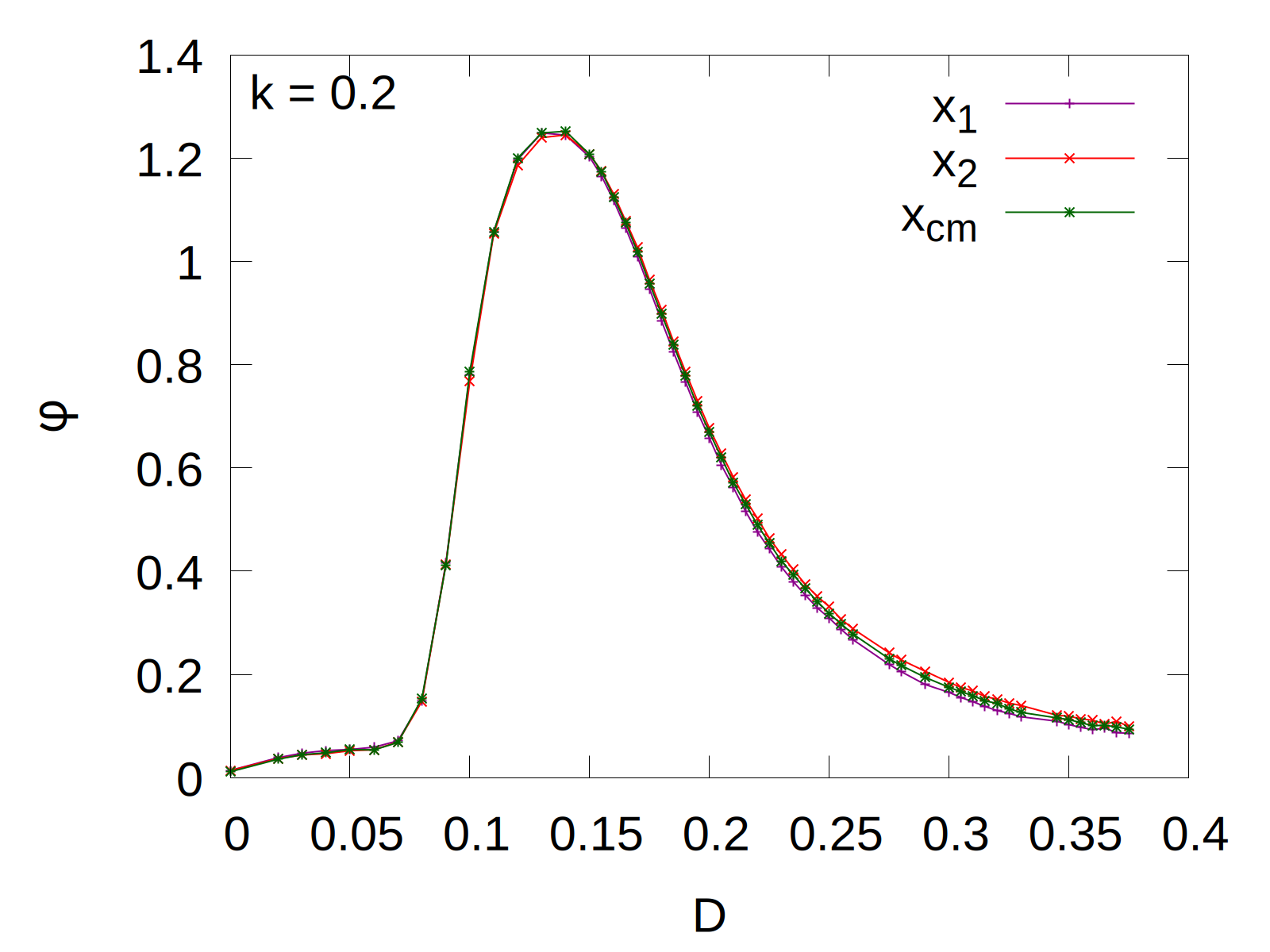}
        \caption{$k=0.2$}
        \label{fig:plimg2}
    \end{subfigure}
    \hfill
    \begin{subfigure}[b]{0.4\textwidth}
        \centering
        \includegraphics[width=\textwidth]{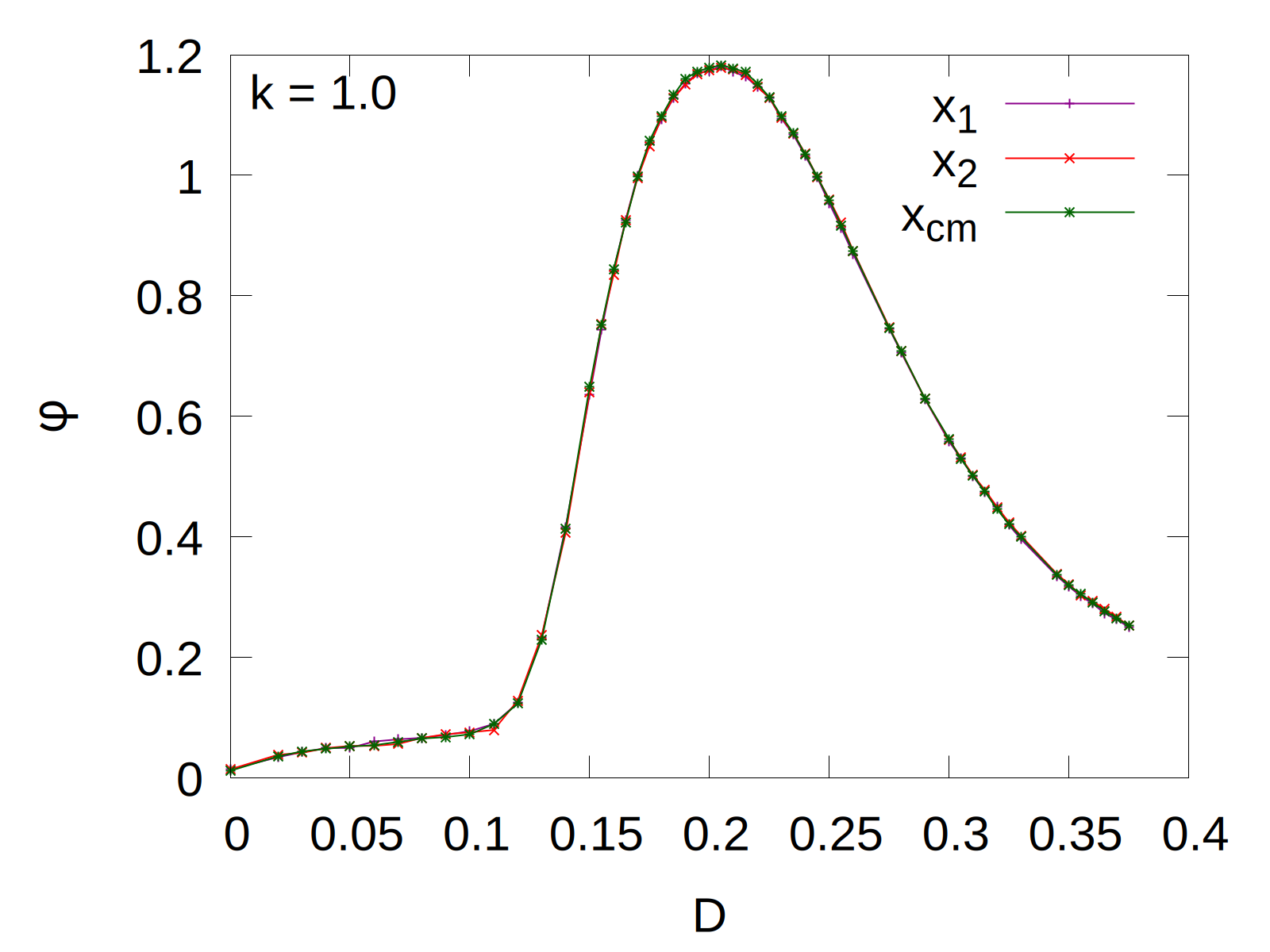}
        \caption{$k=1.0$}
        \label{fig:plimg3}
    \end{subfigure}
    \caption{Phase lag $({\phi})$ vs noise intensity $D$ for different regimes of the dimer i.e. (a) $k=0.05$, (b) $k=0.2$ and (c) $k=1.0$.}
    \label{fig:plag}
\end{figure}

In the context of stochastic resonance, phase lag($\phi$) refers to the delay between the response of the system and the periodic drive input to the system. In other words, it is the lag between the inter-well resonant motion and the input periodic drive. Earlier studies for a single particle \cite{gammaitoni1998stochastic, heinsalu2009stochastic} show the response of a bistable system to a periodic input of small amplitude is related by ,
\begin{equation}
    \langle x(t)\rangle_{as}=x_A(D)\cos(\Omega t-{\phi}(D))
\end{equation}
where ${\phi}$ is the phase lag between the periodic input and the response of the system and $x_A$ is the amplitude. Fig. \ref{fig:plag} shows the variation of phase lag with noise intensity for the 
trajectories of $x_1$, $x_2$ and $x_{cm}$.
The phase lag is calculated from the hysteresis loops and is measured here in radians. The plots suggest that the phase lag follows a non-monotonic behavior like in the case of a single particle \cite{dykman1992phase,gammaitoni1998stochastic,iwai2001study,heinsalu2009stochastic}. For small values of noise intensity, the dimer stays in either of the wells without making a full or even a partial transition across the barrier. The dimer essentially undergoes intra-well motion in this regime. This leads to small values of the phase lag. The phase lag takes a maximum when the temperature becomes close to $D_{SR}$. The maximum values of phase lag in the vicinity of $D_{SR}$ imply that, on an average, the dimer is able to perform a well-synchronized motion, i.e., the temperature is not strong enough to overcome the coupling induced synchronized movement of the dimer. This trend is similar in all the three coupling regimes studied, with the region of temperature values (where the inter-well transitions start making significant contributions to the phase lag), increase to larger values of temperature as the coupling becomes stronger. For much higher temperatures, the monomers often fail to make simultaneous transitions across the barrier. This leads to the occurrence of a difference in the phase lag of the monomers for weak coupling strengths at large temperatures. 

\begin{figure}[H]
    \centering
    \includegraphics[width=0.55\textwidth]{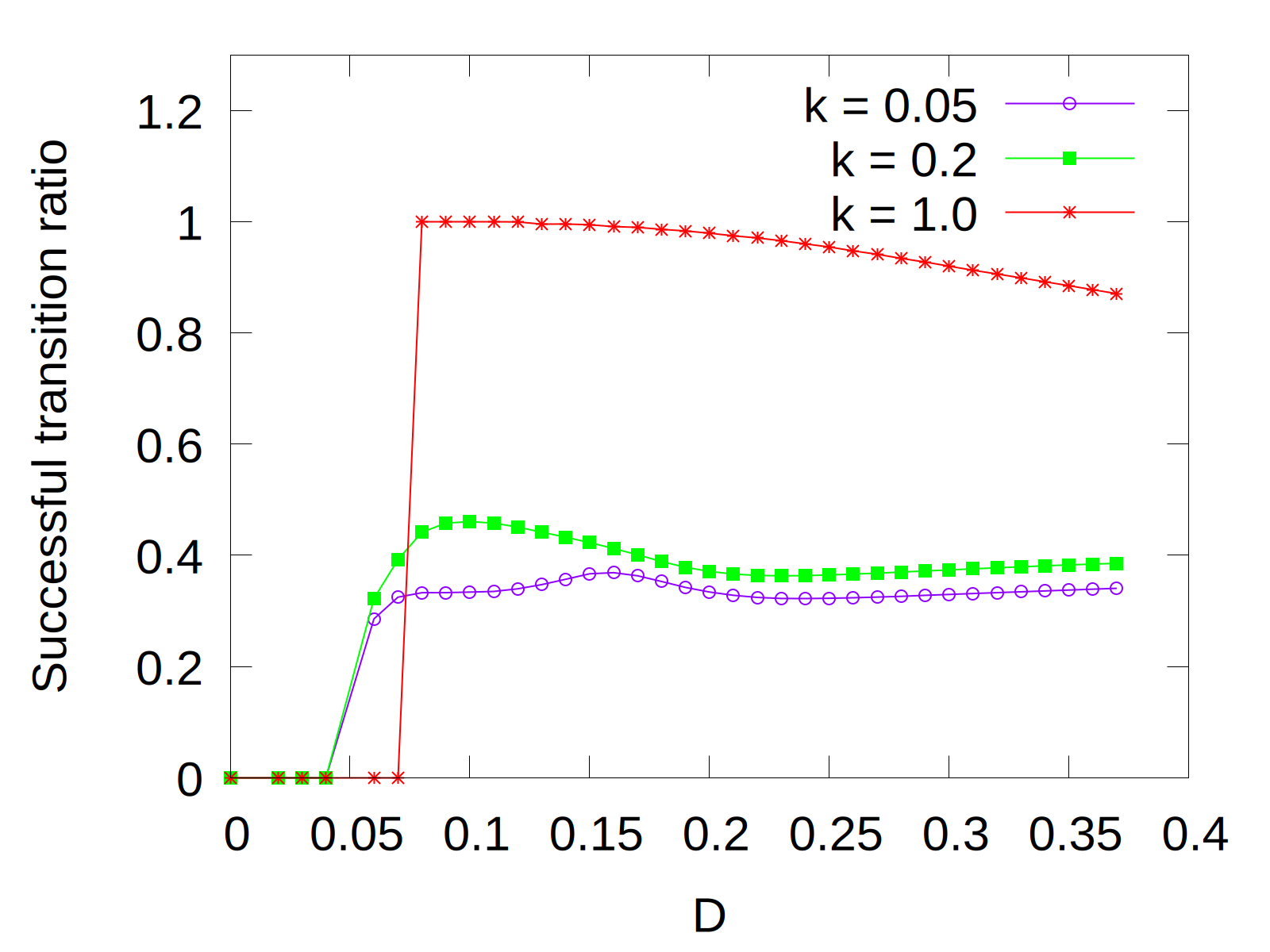}
    \caption{Successful transition ratio vs D for different values of k.}
    \label{fig:ratio}
\end{figure}

In our study, we have defined a coupling induced synchronized movement of the dimer. The synchronized movement here refers to one of the monomers consequently following the motion of the other across the barrier. The escape rate of the dimer is defined accordingly as the dimer is only considered to have made a successful transition across the barrier if both the monomers have crossed a certain threshold position (i.e. $x_c = \pm 0.85$) of the other well. In order to support our findings we have defined another quantity that measures the total number of successful transitions of the dimer across the barrier out of the total transition attempts. We call this quantity the successful transition ratio.
 Suppose the whole dimer starts in the left well. The monomer close to the barrier initiates the process by making a transition over the barrier and crossing the threshold $+x_c$. Now, this monomer can either go back to the left well leading to a failed transition or can stay in the right well till the second monomer also makes a transition to it. The latter will be considered as a successful transition. Fig. \ref{fig:ratio} shows the successful transition ratio for the three coupling regimes of the dimer. 

 The plots suggest that for small values of temperature there are no transitions at all, since the thermal energy is too small to initiate any crossing events. As $D$ increases, the ratio increases gradually in the region where the temperature fluctuations are strong enough to match the restoring forces due to the harmonic part of the interaction potential. This is the region around the first peak in the plots. For a soft dimer, the contributions due to the coupling strength are easily matched by that from the smaller values of noise. This is shown by the small flat region in the plot corresponding to $k=0.05$ around $D=0.1$. On increasing the temperature the successful transition ratio  contains another peak at a value of temperature which is close to the $D_{SR}$ value. This peak corresponds to the contributions of SR in enhancing the successful transitions for small values of coupling strength where the effects of SR are more prominent as the coupling is very weak. In the case of stronger coupling, the contribution due to SR is dominated by that of the coupling induced synchronized movement of the dimer. Hence, the second peak is absent. In the case of strong coupling strengths, for small temperatures, the total number of transitions attempted per trajectory itself is very small, i.e., usually below $30$. The ratio, however, stays close to 1, as in most cases a transition once initiated, i.e., one of the monomers crossing over the barrier, pulls the other monomer with it. It should be noted here that the number of transition attempts at lower values of temperature is maximum for a soft dimer $(\sim 10^3)$ and gets reduced ($\sim 800$) for intermediate values of k. The increase in the ratio suggests a positive contribution of the coupling strength on the transition dynamics of the dimer in the sense of coupling induced synchronized movement but at the same time leading to a reduction in the number of transitions.

\subsection{\label{sec:IE}Average work done per period \protect}
\begin{figure}[H]
    \centering
    \begin{subfigure}[b]{0.47\textwidth}
        \centering
        \includegraphics[width=\textwidth]{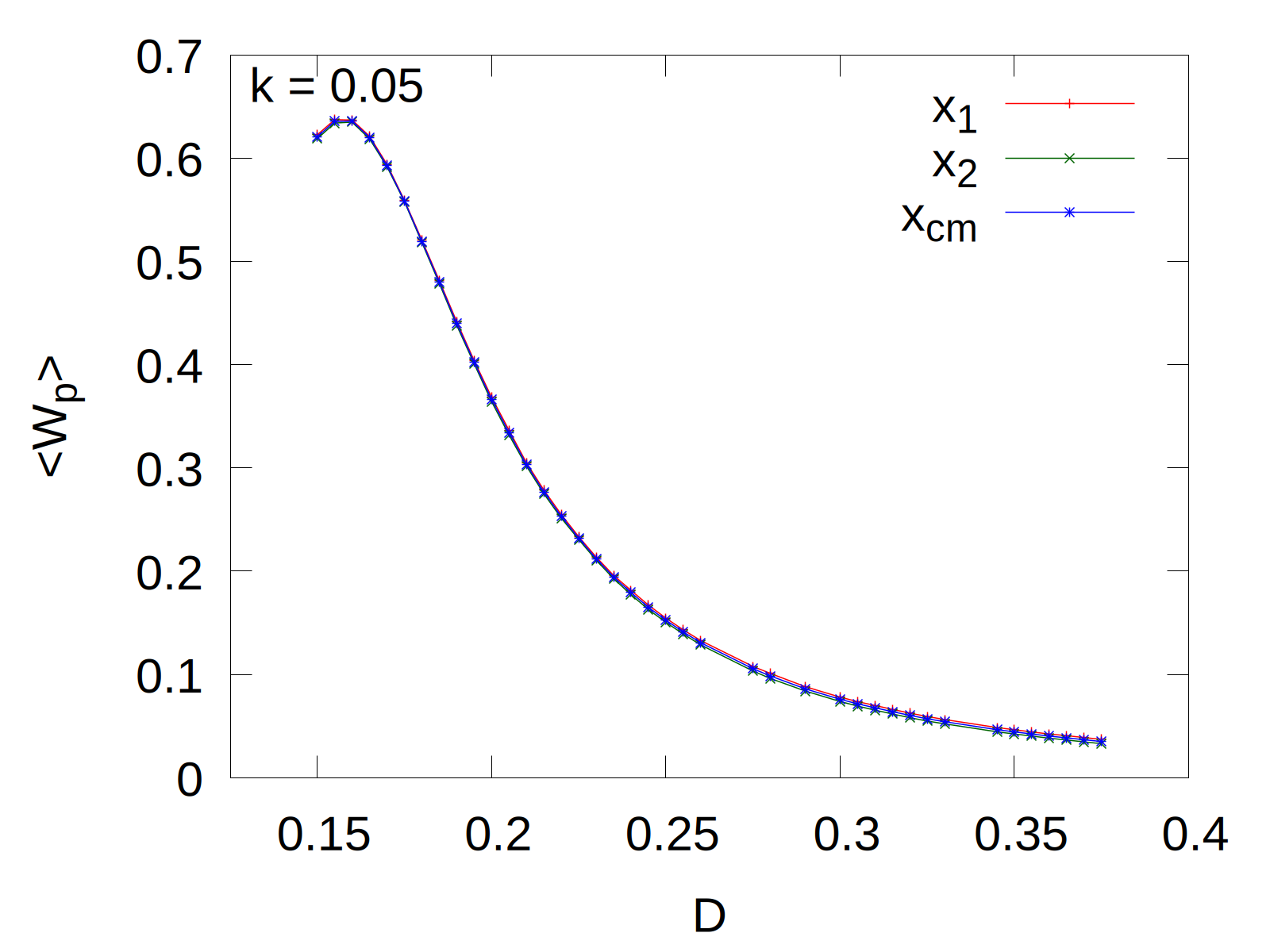}
        \caption{$k=0.05$}
        \label{fig:IEimg1}
    \end{subfigure}
    \hfill
    \begin{subfigure}[b]{0.47\textwidth}
        \centering
        \includegraphics[width=\textwidth]{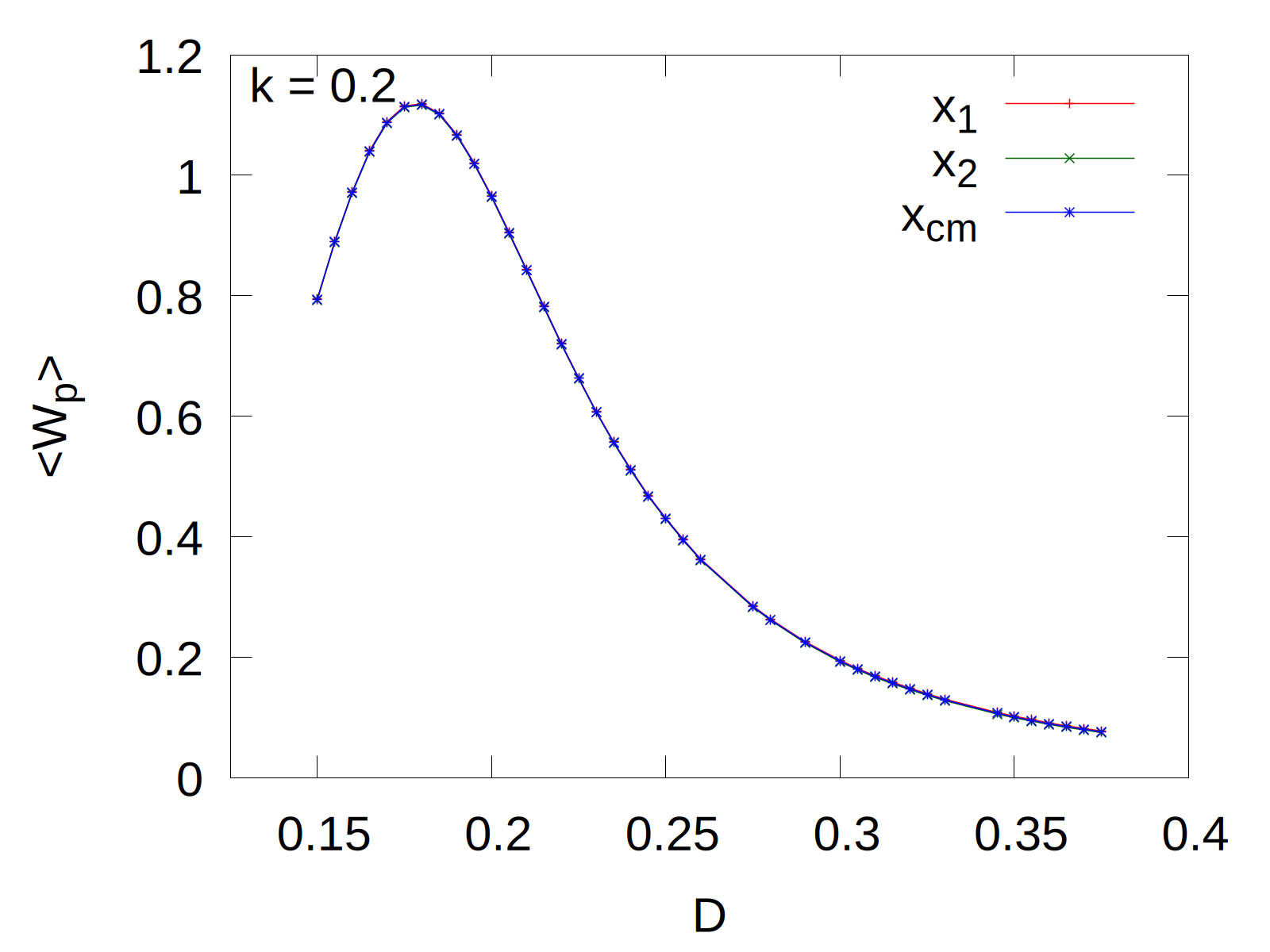}
        \caption{$k=0.2$}
        \label{fig:IEimg2}
    \end{subfigure}
    \hfill
    \begin{subfigure}[b]{0.47\textwidth}
        \centering
        \includegraphics[width=\textwidth]{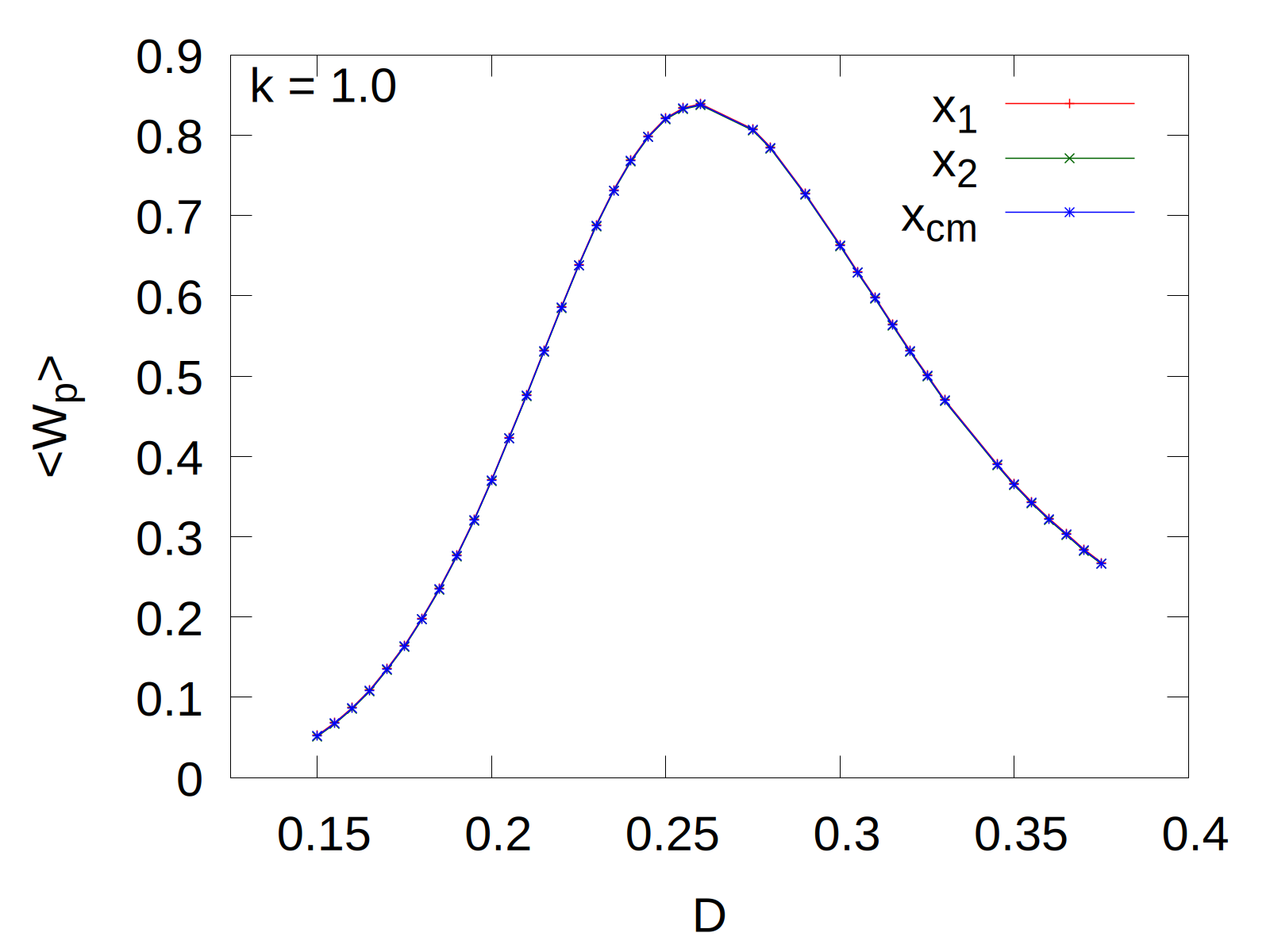}
        \caption{$k=1.0$}
        \label{fig:IEimg3}
    \end{subfigure}
    \caption{Average work done per period of the drive ($W_p$) vs noise strength (D) for the 3 regimes of the dimer i.e. (a) $k=0.05$, (b) $k=0.2$ and (c) $k=1.0$. The green line is for the $W_p$ of center of mass of the dimer $(x_{cm})$, purple and red lines are for the $W_p$ of $x_1$ and $x_2$ respectively.}
    \label{fig:IEthree_images}
\end{figure}

\begin{figure}[H]
    \centering
    \includegraphics[width=0.55\textwidth]{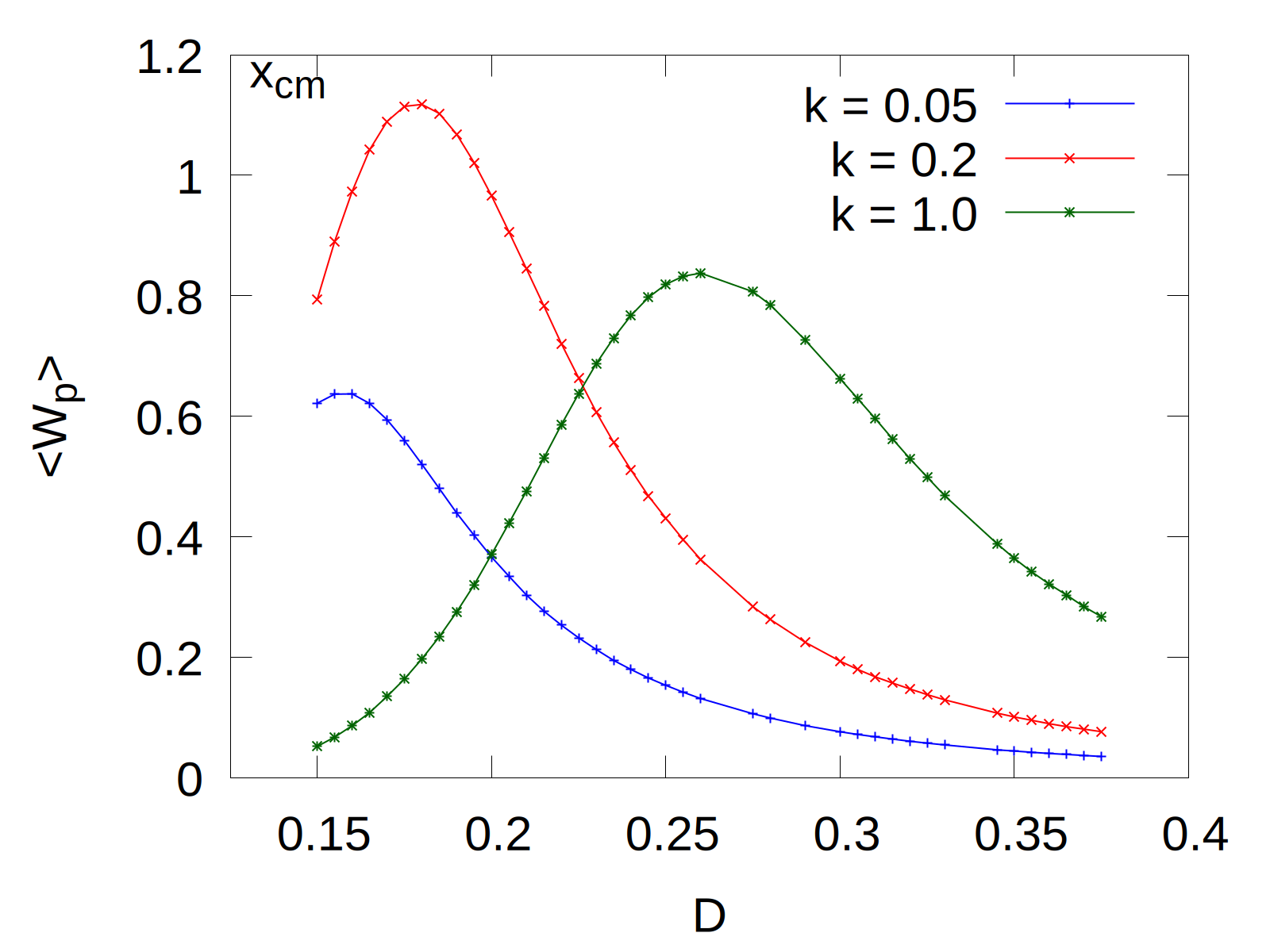}
    \caption{Comparison of $W_p$ of $x_{cm}$ vs D for different values of k.}
    \label{fig:IEvsdcmp}
\end{figure}

The average work done on the system per period $(W_p)$ by the external drive having a frequency $ \Omega=0.018$ is investigated. Fig. \ref{fig:IEthree_images} shows how the ensemble averaged input energy varies with the noise strength for the three different regimes of coupling. The plots for $x_1$, $x_2$ and $x_{cm}$ almost overlap to a good extent. The position of the peaks move towards higher values of $D$ as we go from a soft to a hard dimer. 
The overall behavior observed in all the three plots is similar to that obtained for a single particle \cite{saikia2007work,dan2005bona,iwai2001study}. For small values of noise strength, the dimer does not get enough time to absorb sufficient energy from the fluctuations to surmount the barrier even once in a single cycle of the external drive, whereas for large values of noise strength, the dimer makes several transitions across the barrier in a single cycle of the external drive. In terms of the average work done per period by the external drive on the system, consider first the case when the rate at which the dimer makes a transition across the barrier is less than twice the frequency of the external drive $(D<D_{SR})$. Suppose the dimer's center of mass is near $x=-1$. For the first half period, the external force will act in the positive $x-$direction. The dimer's effective displacement due to the harmonic, LJ and bistable forces together with the random fluctuations due to the noise can very well be in either directions i.e. in the positive $x-$direction or in the negative $x-$direction. As long as the dimer stays in the same well these displacements will be small and hence the work done by the external force on the system will also be small. The work done by the external force can hence be either positive or negative, depending upon the directions of displacement and the external force.
   In the second half period the force will start acting in the negative $x-$ direction. If the dimer makes a transition from the left to the right well in such a condition, the displacement will be in the positive direction of $x$ while the external force will be acting opposite to it. This would lead to negative values of $W_p$. During a transition, the work done by the external force (both positive and negative) are considerably more significant than the work done during intra-well motion. Since there is a mismatch between the two mentioned frequencies, the total work done per period averaged over several cycles and trajectories will be smaller when compared to the optimum case. When the temperature value is close to $D_{SR}$, the transitions between wells are in coherence with the external drive, i.e. the transition in the positive direction occurs when the force is acting in the positive direction and vice-versa. In both forward and backward transitions, since the force and displacement act along the same direction, the work done by the force is positive. The total $W_p$ per period will essentially increase leading to maximum values for optimum temperatures. 
For large temperatures, the dimer will make transitions more frequently leading to multiple transitions in a single cycle of the external drive. In such a case, the inter-well displacement will change its direction multiple times during each cycle leading to force and displacement having both opposite and same directions. This will again reduce the average values of $W_p$.
 Fig. \ref{fig:IEvsdcmp} shows a comparison between the $W_p$ characterized by the center of mass in the three coupling regimes of the dimer. It is evident from the plot that different values of coupling strengths also influence the average work done per period by the external drive.

\begin{figure}[H]
    \centering
    \includegraphics[width=0.55\textwidth]{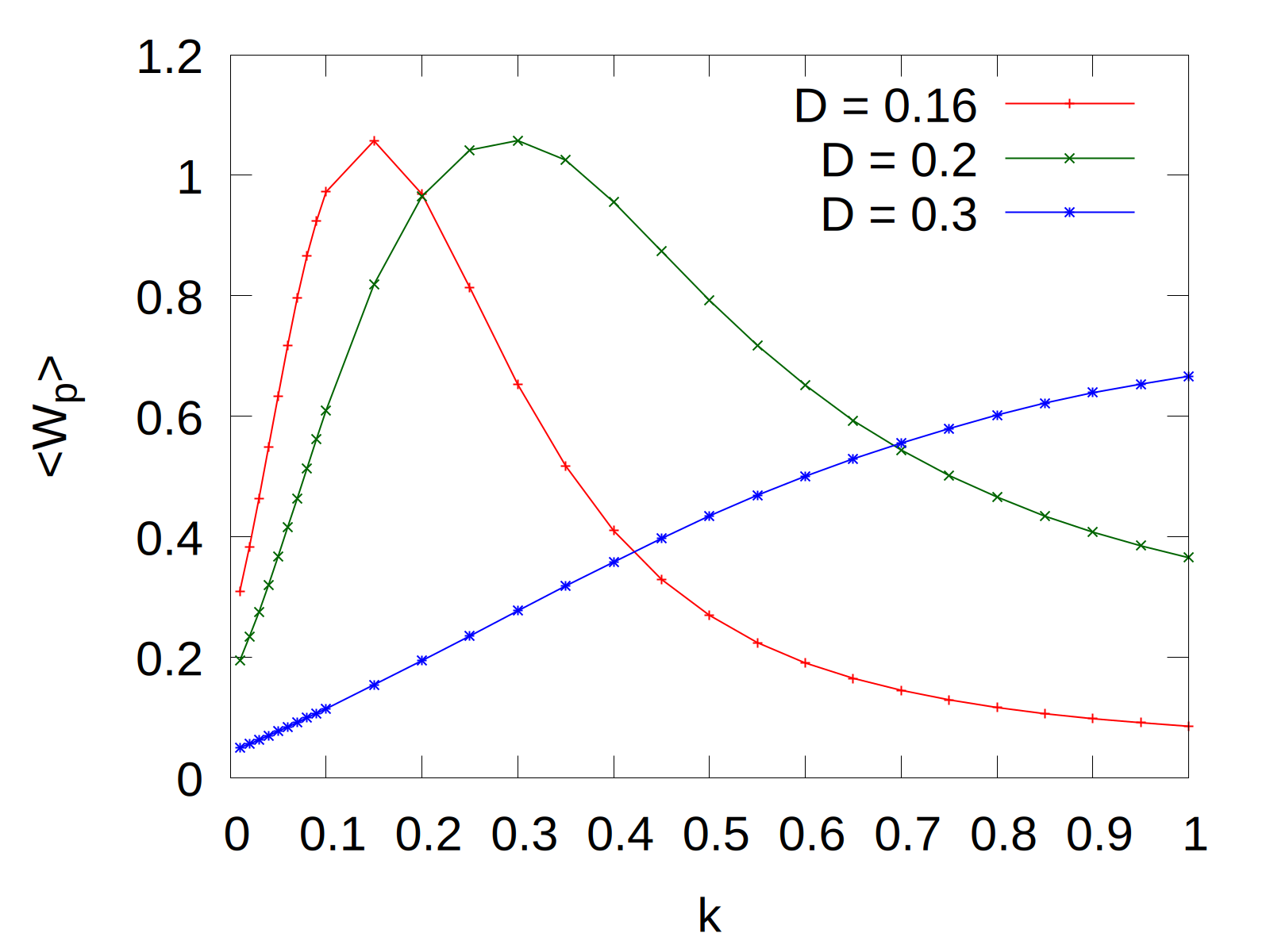}
    \caption{Comparison of $W_p$ of $x_{cm}$ vs k for different values of D.}
    \label{fig:wpvk}
\end{figure}

\begin{figure}[H]
    \centering
    \begin{subfigure}[b]{0.46\textwidth}
        \centering
        \includegraphics[width=\textwidth]{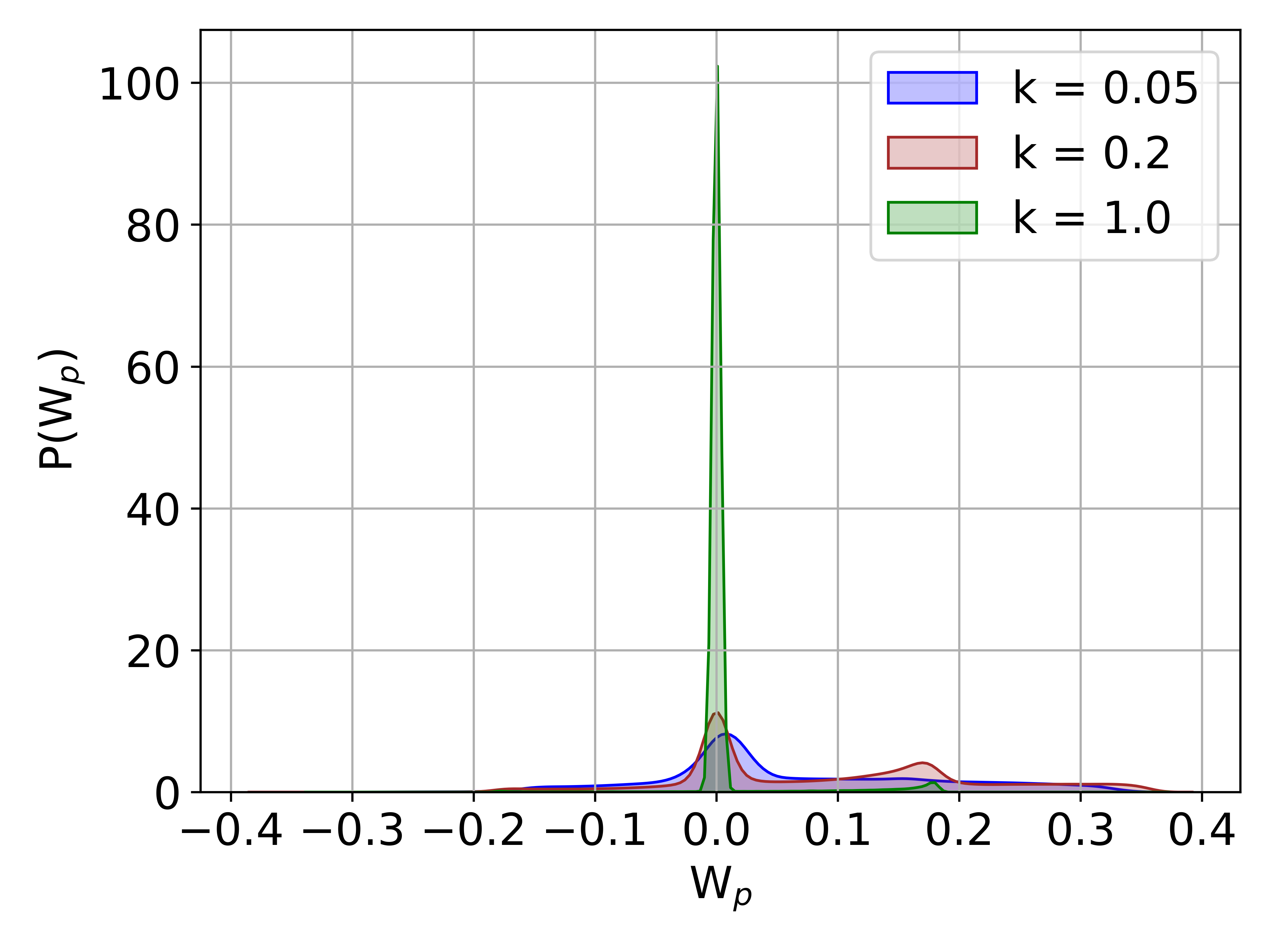}
        \caption{$D=0.155$}
        \label{fig:pdDimg1}
    \end{subfigure}
    \hfill
    \begin{subfigure}[b]{0.46\textwidth}
        \centering
        \includegraphics[width=\textwidth]{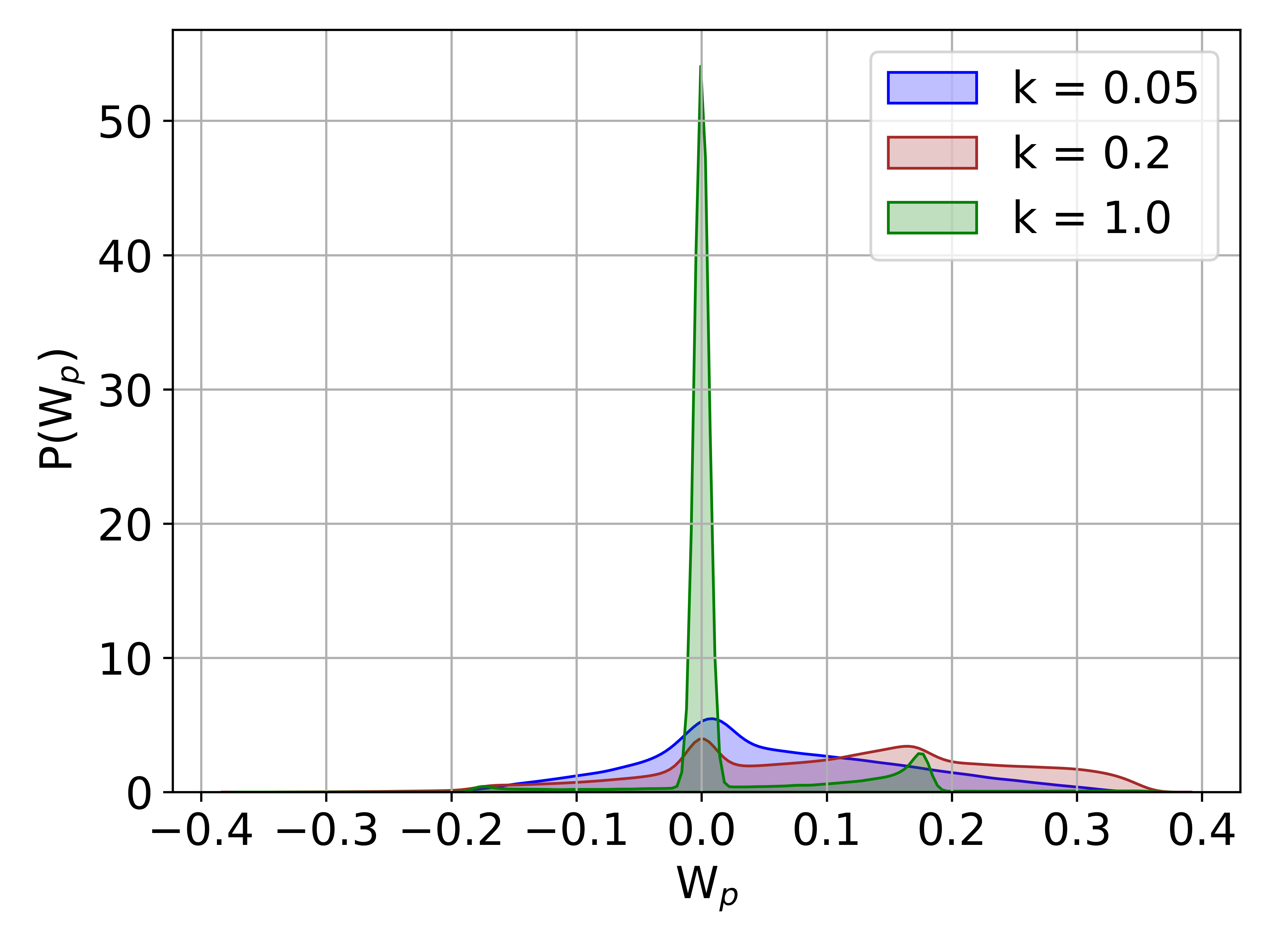}
        \caption{$D=0.18$}
        \label{fig:pdDimg2}
    \end{subfigure}
    \hfill
    \begin{subfigure}[b]{0.46\textwidth}
        \centering
        \includegraphics[width=\textwidth]{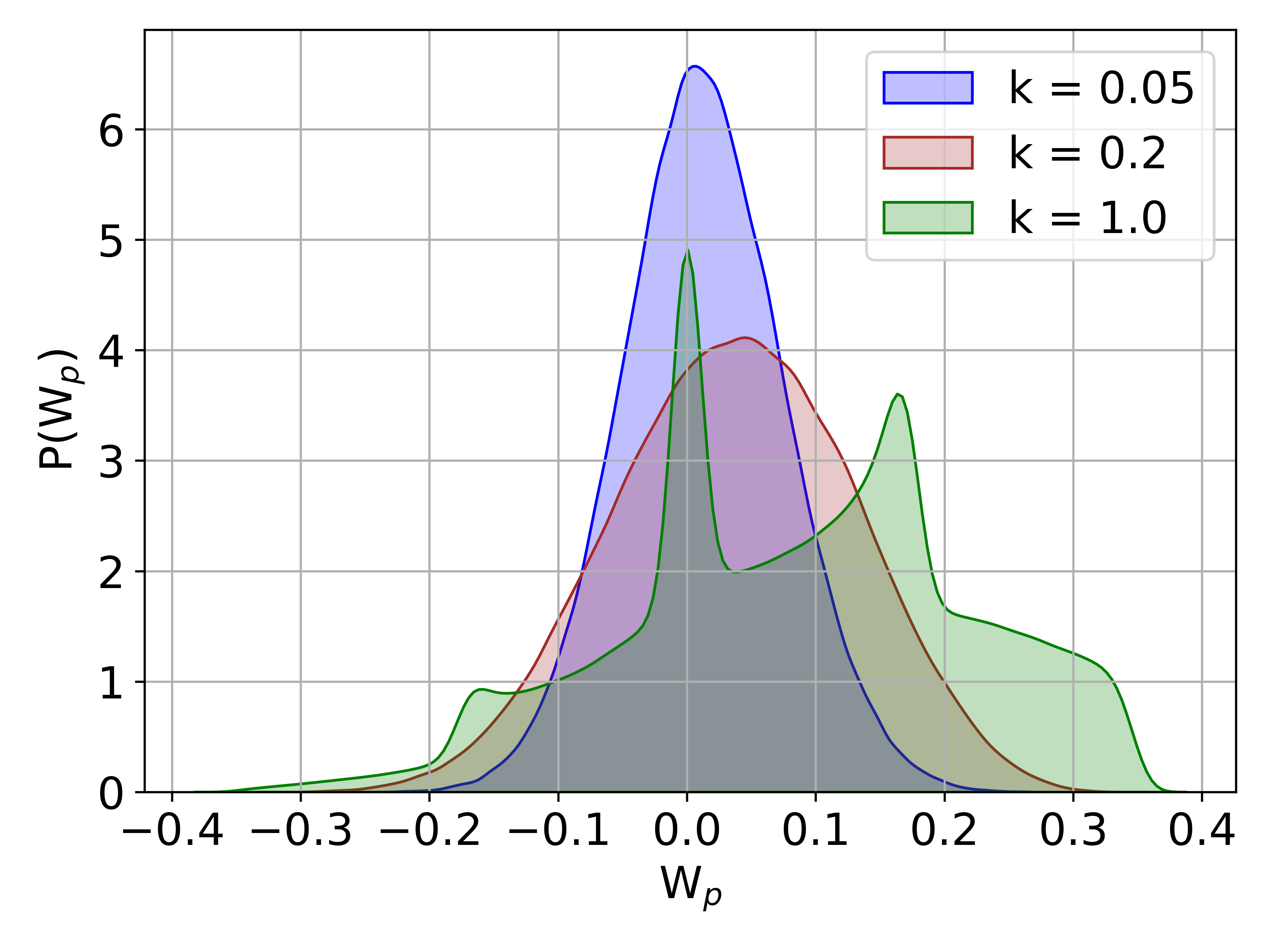}
        \caption{$D=0.26$}
        \label{fig:pdDimg3}
    \end{subfigure}
    \hfill
    \begin{subfigure}[b]{0.46\textwidth}
        \centering
        \includegraphics[width=\textwidth]{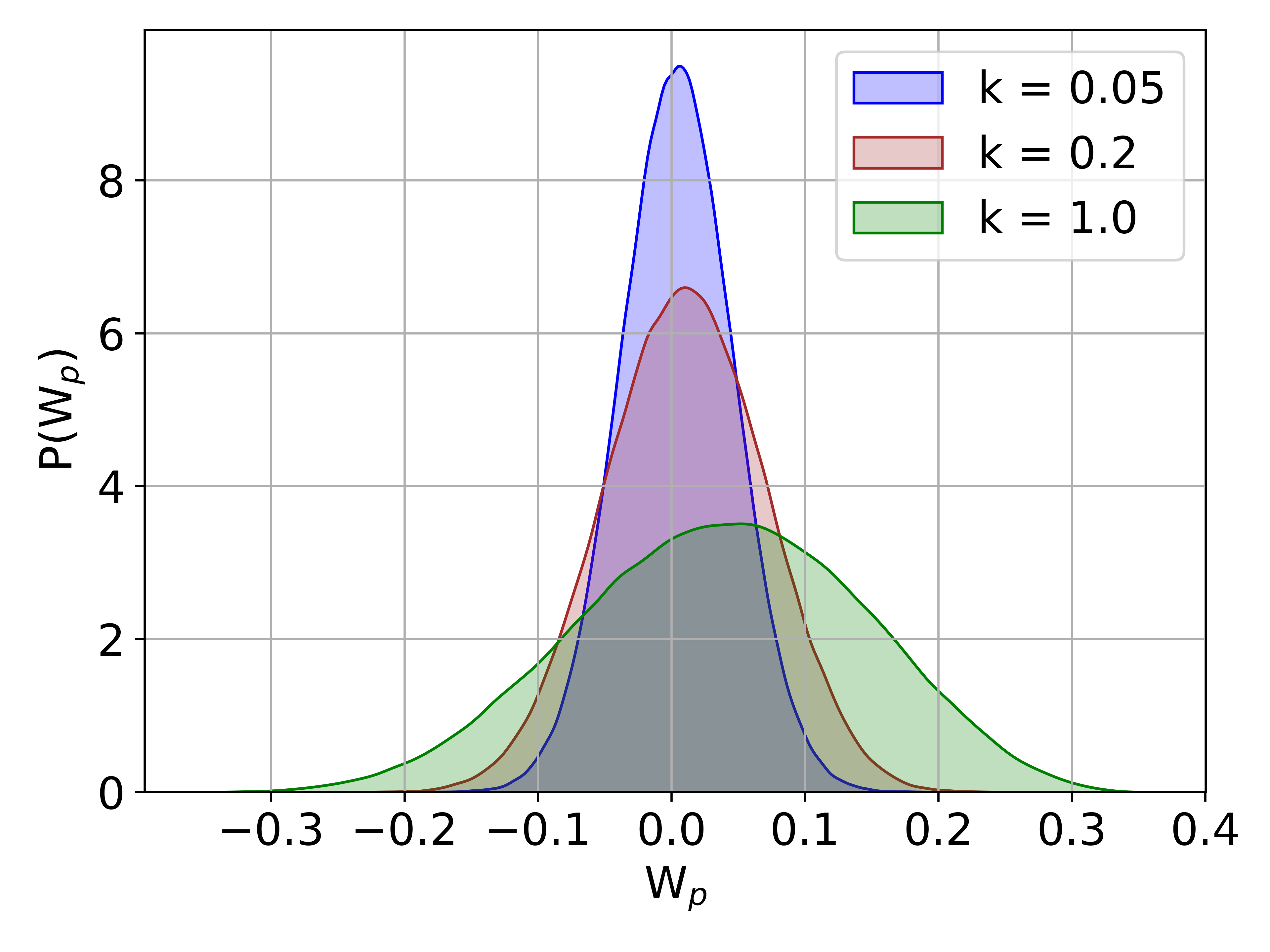}
        \caption{$D=0.35$}
       \label{fig:pdDimg4}
    \end{subfigure}
    
    \caption{Probability distribution of $W_p$ for different noise strengths (a) $D=0.155$, (b) $D=0.18$, (c) $D=0.26$ and (d) $D=0.35$.}
    \label{fig:pdDfour_images}
\end{figure}

 A peaking behavior is observed in Fig. \ref{fig:wpvk} where the variation of $W_p$ is studied with respect to the coupling strength of the dimer for 3 different values of temperature. The work done first increases and then decreases as we increase the coupling strength of the dimer. 
 For small coupling strengths at a fixed temperature, the dimer requires less energy to cross the barrier and hence is able to make several transitions back and forth across the barrier. The dimer gets more than enough time to absorb the energy from the thermal fluctuations and thus does not wait for the most suitable condition depicted in Fig. \ref{fig:suitcnd}, to make a transition. For large values of coupling strength, the dimer fails to acquire enough energy to overcome the barrier even when the barrier height is minimum. This failure can  be attributed to the strong restoring forces that hold both the monomers together and prevent the dimer from making even a single transition per period of the external drive. For the intermediate values of coupling strength, the dimer gets just enough time to acquire sufficient energy to make a transition when the barrier height is minimum. Thus, the dimer is able to make a jump to the other well once every half a period of the external drive. This enhances the response of the system and leads to a peak in the $W_p$ versus $k$ plot. 
A better comparison is done using the probability distribution of the input energy at fixed temperatures.
Fig. \ref{fig:pdDfour_images} shows the probability distribution of the average work done on the system of a dimer by the external periodic force. We have compared the distributions for the three coupling regimes at different noise strengths. Probability distributions for input energy per period have been studied for the system of a single particle in a bi-stable potential \cite{saikia2007work} and in periodic potentials \cite{saikia2011stochastic,reenbohn2012periodically,saikia2014role}. Earlier studies show the multipeak structure in the distribution of $W_p$ to be an indicator of SR. The most asymmetric distribution corresponds to the value of parameters for which the system exhibits SR. For small values of noise strengths shown in Fig. \ref{fig:pdDimg1} and \ref{fig:pdDimg2}, we observe a large peak around $W_p = 0$ for a hard dimer. The strong restoring forces in a hard dimer restrict its motion around the bottom of a single well and it does not get enough time to absorb sufficient energy from the thermal fluctuations to make even a single transition in one period of the drive. This intra-well motion of the hard dimer contributes to the sharp peaks in the plots. The height of this peak decreases and gradually leads to a multi-peak structure with the onset of inter-well transitions.
\noindent  The small second peak around $W_p\approx0.2$ is associated with inter-well motion. On increasing the temperature, the dynamics is dominated by a number of frequent transitions across the barrier in a single cycle of the external drive. The total work done per period can be considered as the sum of small independent increments in it. On applying the central limit theorem, we can expect the peak to be Gaussian in this limit.
This discussion supports our previous findings, giving a better picture of how the dynamics of the dimer are influenced by the temperature and the coupling.

\section{Conclusions and discussion}\label{sec4}
In this study, SR has been investigated in an overdamped system of a dimer (two coupled oscillators) confined to a bistable potential in three distinct coupling regimes. The interaction between the substructures was defined by the sum of harmonic and the LJ potential. First we have analyzed the HLA in order to quantify how energy is dissipated in the system and its variation with the coupling strength and the temperature. Next, we have examined the ensemble averaged trajectories of the center of mass of the dimer to describe the collective behavior of the system. The maximum amplitude of oscillation presented an increase in the amplitude for intermediate values of coupling. This is the regime where the coupling is sufficient enough for the synchronized movement of the dimer but is not that strong to reduce its flexibility. We have used the hysteresis loops to calculate the phase lag between the dimer's response and the periodic input signal. On comparing the phase lag of the individual monomers with that of the center of mass of the dimer, we found that for weak coupling strengths and large temperatures, the monomers loose their synchronization, leading to a difference in the corresponding phase lag plots. An important contribution of this work is the introduction of successful transition ratio, a metric defined to study the coupling induced synchronized movement of the dimer. It highlights the regions where SR actually contributes in increasing the successful transition ratio. The contribution of SR towards more coherent or synchronized transitions was only significant for weak coupling strengths, where the initial flat region of the curve corresponded to the balancing of fluctuations due to noise and the coupling effects while the peak corresponded to the contribution of SR in increasing the successful transition ratio. For intermediate and large coupling strengths the contribution due to SR was absent. In terms of input energy, the dimer possessed sufficient energy to overcome the coupling effects even at small temperatures. This lead to less efficient transfer of energy from the external signal to the system. When the coupling was increased, the temperatures were just correct for the dimer to exhibit SR and are also not too large so as to completely overcome the synchronization brought about by the coupling. This lead to more coherent transitions of the monomers leading to an increase in the input energy transfer. For a hard dimer the flexibility was lost putting a constraint on the motion of the dimer. From this study we have examined how the coupling strength and the temperature cooperate and influence the resonance phenomena. Our study can contribute to a broader understanding in the field of SR in coupled systems. These principles can help in the application based studies of several coupled systems working in noisy environments especially in the field of energy harvesters. 

\bibliography{manu}

\end{document}